\documentclass[11pt]{article}

\usepackage[top=1in,bottom=1in,left=1in,right=1in]{geometry}
\usepackage{natbib}
\usepackage[unicode=true,bookmarks=true,bookmarksnumbered=false,bookmarksopen=false,breaklinks=false,pdfborder={0 0 1},backref=false,colorlinks=true]{hyperref}
\hypersetup{citecolor=blue}
\usepackage{url}

\usepackage{amsmath}
\usepackage{mathrsfs}
\usepackage{amssymb}
\usepackage{amsthm}
\usepackage[normalem]{ulem}

\newtheorem{theorem}{Theorem}

\newtheorem{proposition}{Proposition}

\newtheorem{definition}{Definition}

\usepackage{float}
\usepackage{rotfloat}

\floatstyle{ruled}
\newfloat{algorithm}{tpb}{loa}
\providecommand{\algorithmname}{Algorithm}
\floatname{algorithm}{\protect\algorithmname}

\usepackage{algpseudocode,algorithm}
\algnewcommand\algorithmicinput{\textbf{Input}:}
\algnewcommand\algorithmicoutput{\textbf{Output}:}
\algnewcommand\INPUT{\item[\algorithmicinput]}
\algnewcommand\OUTPUT{\item[\algorithmicoutput]}

\usepackage{graphicx}
\usepackage[caption=false,labelformat=simple]{subfig}

\usepackage{array}
\newcolumntype{L}[1]{>{\raggedright\let\newline\\\arraybackslash\hspace{0pt}}m{#1}}
\newcolumntype{C}[1]{>{\centering\let\newline\\\arraybackslash\hspace{0pt}}m{#1}}
\newcolumntype{R}[1]{>{\raggedleft\let\newline\\\arraybackslash\hspace{0pt}}m{#1}}

\usepackage{rotating}
\usepackage{multirow}

\usepackage{tikz}
\usetikzlibrary{shapes,arrows,backgrounds,calc,positioning,fit,petri,plotmarks}
\usetikzlibrary{arrows}

\usepackage{enumerate}
\usepackage{paralist}

\newcommand*{\affaddr}[1]{#1} 
\newcommand*{\affmark}[1][*]{\textsuperscript{#1}}

\global\long\def\bX{\mathbf{X}}
\global\long\def\bx{\mathbf{x}}
\global\long\def\bY{\mathbf{Y}}

\global\long\def\bz{\mathbf{z}}

\global\long\def\bD{\mathbf{D}}
\global\long\def\bA{\mathbf{A}}

\global\long\def\bgamma{\boldsymbol{\gamma}}
\global\long\def\bbeta{\boldsymbol{\beta}}

\global\long\def\bepsilon{\boldsymbol{\epsilon}}
\global\long\def\bvarepsilon{\boldsymbol{\varepsilon}}
\global\long\def\bOmega{\boldsymbol{\Omega}}
\global\long\def\btheta{\boldsymbol{\theta}}

\newcommand{\indep}{\rotatebox[origin=c]{90}{$\models$}}

\title{Identifying brain hierarchical structures associated with Alzheimer's disease using a regularized regression method with tree predictors}

\author{%
    Yi Zhao\affmark[1], Bingkai Wang\affmark[2], Chin-Fu Liu\affmark[3], Andreia V. Faria\affmark[4], Michael I. Miller\affmark[3], \\ 
    Brian S. Caffo\affmark[2], Xi Luo\affmark[5], and
    for the Alzheimer's Disease Neuroimaging Initiative\footnote{Data used in preparation of this article were obtained from the Alzheimer's Disease Neuroimaging Initiative (ADNI) database (\url{adni.loni.usc.edu}). As such, the investigators within the ADNI contributed to the design and implementation of ADNI and/or provided data but did not participate in analysis or writing of this report. A complete list of ADNI investigators can be found at: \url{http://adni.loni.usc.edu/wp-content/uploads/how_to_apply/ADNI_Acknowledgement_List.pdf}} \\
    \affaddr{\affmark[1]Department of Biostatistics and Health Data Science, Indiana University School of Medicine} \\
    \affaddr{\affmark[2]Department of Biostatistics, Johns Hopkins Bloomberg School of Public Health} \\
    \affaddr{\affmark[3]Center for Imaging Science, Biomedical Engineering, Johns Hopkins University} \\
    \affaddr{\affmark[4]Department of Radiology, Johns Hopkins University School of Medicine} \\
    \affaddr{\affmark[5]Department of Biostatistics and Data Science,\\  The University of Texas 
Health Science Center at Houston} \\
}

\date{}

\providecommand{\keywords}[1]
{
  {\small	
  \textbf{Keywords:} #1}
}

\begin{document}

\maketitle

\thispagestyle{empty}

\begin{abstract}
Brain segmentation at different levels is generally represented as hierarchical trees. Brain regional atrophy at specific levels was found to be marginally associated with Alzheimer's disease outcomes. 
In this study, we propose an $\ell_{1}$-type regularization for predictors that follow a hierarchical tree structure. Considering a tree as a directed acyclic graph, we interpret the model parameters from a path analysis perspective. Under this concept, the proposed penalty regulates the total effect of each predictor on the outcome. With regularity conditions, it is shown that under the proposed regularization, the estimator of the model coefficient is consistent in $\ell_{2}$-norm and the model selection is also consistent. By applying to a brain structural magnetic resonance imaging dataset acquired from the Alzheimer's Disease Neuroimaging Initiative, the proposed approach identifies brain regions where atrophy in these regions demonstrates the declination in memory. With regularization on the total effects, the findings suggest that the impact of atrophy on memory deficits is localized from small brain regions but at various levels of brain segmentation.
\end{abstract}
\keywords{Generalized Lasso; Path analysis; Tree Data}




\section{Introduction}

In the problem of linear regression with high-dimensional data, $\ell_{1}$-regularization and its variations are ubiquitously studied. Well-known examples include the lasso~\citep{tibshirani1996regression}, the elastic net~\citep{zou2005regularization}, the adaptive lasso~\citep{zou2006adaptive}, the fused lasso~\citep{tibshirani2005sparsity}, the group lasso~\citep{yuan2007model}, the generalized lasso~\citep{tibshirani2011solution}, and many others. Among these, the fused lasso, the group lasso and the generalized lasso accommodate certain structural information in the design matrix to achieve desired profiles of the model coefficients. In this study, we consider a scenario where the predictors possess a hierarchical tree structure. Treating the hierarchical tree as a directed acyclic graph, we interpret the model parameters from a path analysis perspective and propose a lasso-type penalty that regulates the total effect of the predictor on the outcome.

\begin{figure}
  \begin{center}
    \includegraphics[width=\textwidth]{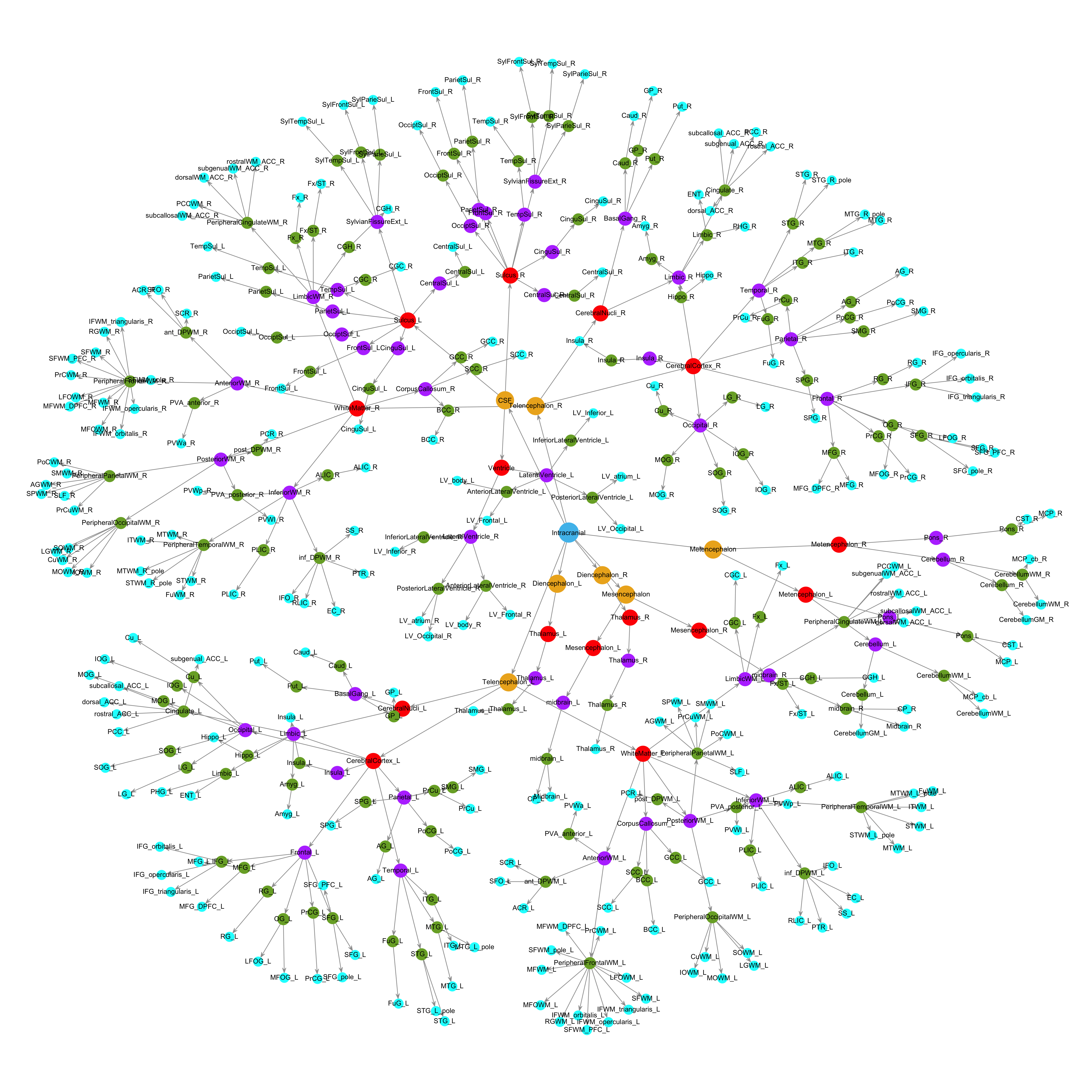}

    \includegraphics[width=0.4\textwidth]{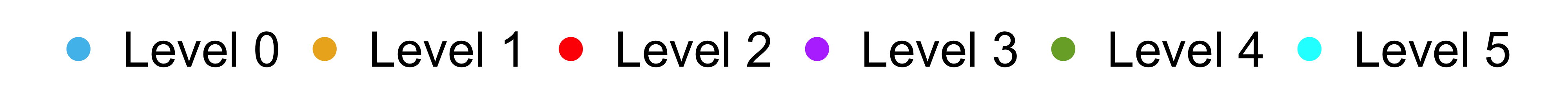}
  \end{center}
  \caption{\label{fig:brain_parcel}The hierarchical tree structure of the brain segmentation. Brain regions are colored by level.}
\end{figure}

This is motivated by structural magnetic resonance imaging (sMRI) studies. With the availability of high-quality 3D images, it is now possible to measure brain morphometry and investigate associations with mental disorders. For example, in the study of Alzheimer's Disease (AD), sMRI is considered as a direct reflection of the density of neurofibrillary tangles, an established pathological hallmark of AD. It captures atrophy in gray matter due to the loss of neurons, synapses, and dendritic de-arborization (atrophy in white matter due to the loss of structural integrity of white matter fiber tracts presumably a consequence of demyelination and dying back of axonal processes) and an ex vacuo (increases in the volume of the cerebralspinal fluid (CSF) caused by the loss of encephalic volume). Thus, measurements acquired from sMRI have been widely used to identify (regional) markers of AD~\citep{vemuri2010role}. In order to extract regional structural data, anatomical brain segmentation is generally applied. Multi-atlas segmentation (MAS) is a popular approach, which has the advantage of coordinating representations from multiple segmented atlases and correcting errors through a label fusion process. Several MAS approaches offer hierarchical segmentations at various granularity levels~\citep{wu2015hierarchical,doshi2016muse,mori2016mricloud}. 
\citet{mori2016mricloud} introduced a segmentation that has a five-level hierarchical structure, starting from a coarse segmentation (Level 1) into major areas (telencephalon, diencephalon, metencephalon, mesencephalon, and CSF) to a fine segmentation (Level 5) defining hundreds of structures, as small as gyri and deep nucleae.
The hierarchical tree structure of this multi-level segmentation is presented in Figure~\ref{fig:brain_parcel}. When studying the association with AD symptoms, such as memory decline, the granularity level of the region of interests (ROIs) that play a role may diverge across the brain. For example, the hippocampus, which is a Level-4 region in the segmentation, is a consistently identified brain region related to memory deficits in all stages of AD~\citep{pini2016brain}. Another well-known marker area is the entorhinal cortex~\citep{pini2016brain}, a Level-5 region. Both the hippocampus and entorhinal cortex are part of the limbic system, which is a Level-3 ROI. Thus, it is beneficial to include data extracted from all levels of segmentation into feature selection and, in the meantime, taking the hierarchical structure and data dependence into consideration.

Considering the tree structure as a directed acyclic graph, we introduce an $\ell_{1}$-type regularization, which incorporates the structural information by taking the influence matrix of the predictors as the penalty matrix. The influence matrix can be either obtained from the weighted adjacency matrix or estimated from the data following the hierarchical structural specification. Using a path diagram, we demonstrate that this is equivalent to regularizing the total effect of each predictor on the outcome. 

The rest of the paper is organized as the following. In Section~\ref{sec:method}, we introduce regularization functions for data following a hierarchical tree structure and show that (under regularity conditions) the regularized estimator of the model parameter and the estimated active set are both consistent. Section~\ref{sec:sim} presents the simulation results demonstrating the performance of the proposed regularizations under various scenarios. In Section~\ref{sec:adni}, we apply the proposed approach to the data collected in the Alzheimer's Disease Neuroimaging Initiative (ADNI). The goal is to investigate the association between atrophy in the brain and memory deficits, where the brain volumetric data are extracted from a hierarchical segmentation. Using the proposed regularization function, the identified relevant brain regions are in line with existing literature. In addition, the findings infer local impacts of atrophy on memory decline. Section~\ref{sec:discussion} summarizes the paper with a discussion.

\section{Methods}
\label{sec:method}

\begin{figure}
  \begin{center}
    \subfloat[\label{fig:method_eg1}Example 1]{\includegraphics[width=0.45\textwidth, page=1]{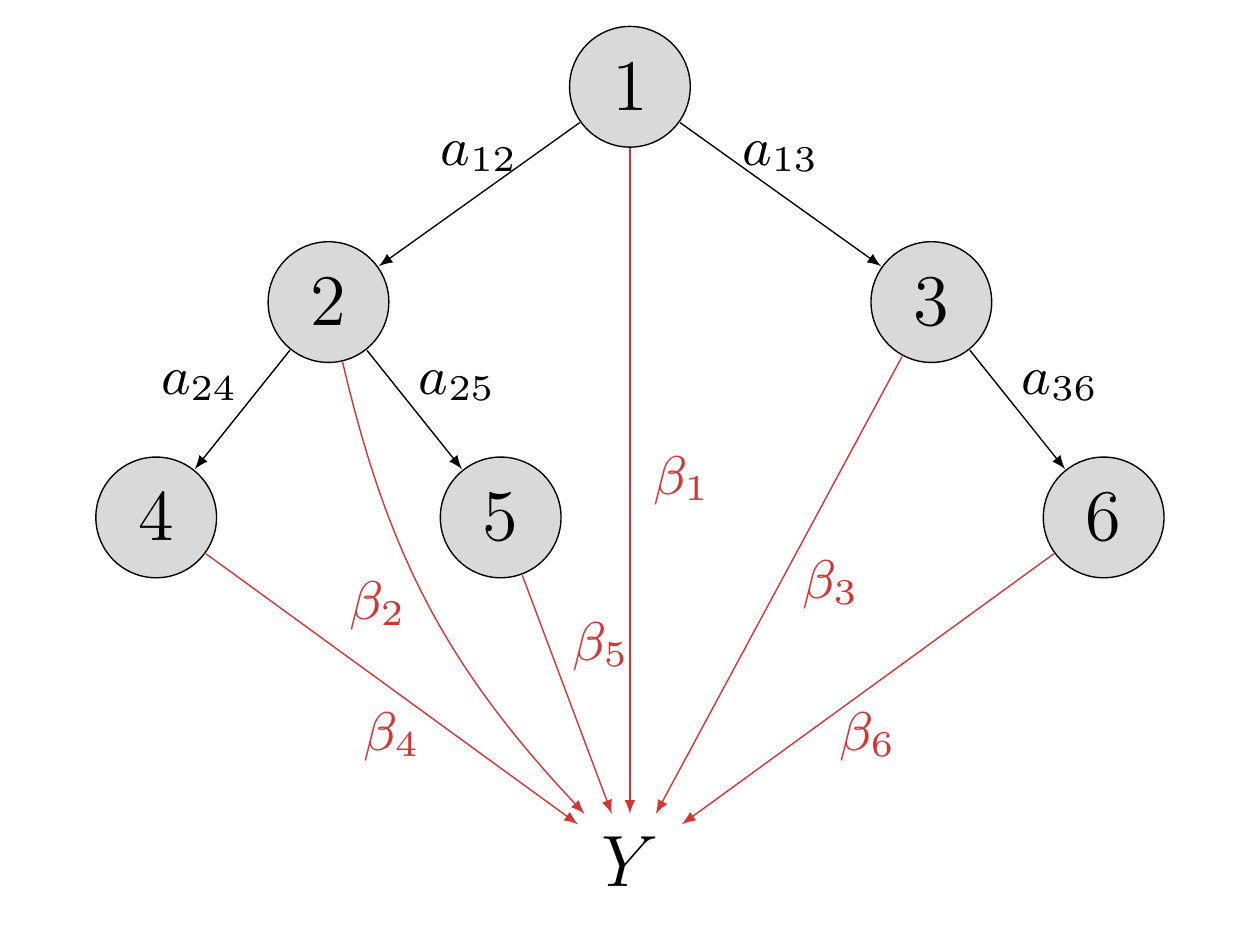}}
    \enskip{}
    \subfloat[\label{fig:method_eg2}Example 2]{\includegraphics[width=0.45\textwidth, page=2]{latex.pdf}}
  \end{center}
  \caption{\label{fig:method_eg}Toy examples with (a) $p=6$ and (b) $p=3$ predictors following a directed hierarchical tree structure.}
\end{figure}

Let $\bY=(y_{1},\dots,y_{n})^\top\in\mathbb{R}^{n}$ denote the outcome vector of $n$ subjects and $\bX=(\bx_{1},\dots,\bx_{n})^\top\in\mathbb{R}^{n\times p}$ the $p$-dimensional design matrix. The following linear regression model is considered:
\begin{equation}\label{eq:model}
  \bY=\bX\bbeta+\bepsilon,
\end{equation}
where $\bbeta\in\mathbb{R}^{p}$ is a $p$-dimensional vector of model coefficients, and elements in $\bepsilon\in\mathbb{R}^{n}$ are the model errors, assumed to be independent and identically distributed following a normal distribution with mean zero and variance $\sigma^{2}$. Different from a regular design matrix, the columns of $\bX$ possess a directed hierarchical tree structure. We propose to represent the tree structure using the following formulation.
\begin{equation}\label{eq:model_X}
  \bX=\bX\bA+\bvarepsilon,
\end{equation}
where $\bA=(a_{ij})\in\mathbb{R}^{p\times p}$ is the weighted \emph{adjacency matrix} of the predictors and $\bvarepsilon\in\mathbb{R}^{n\times p}$ is an error matrix. Without loss of generality, we assume that the predictors are centered with mean zero. Each row of $\bvarepsilon$ follows a $p$-dimensional normal distribution with covariance matrix, $\bOmega$, where $\bOmega$ is a diagonal matrix. In this case, the dependency between the predictors is fully captured by the adjacency matrix. Since a directed hierarchical tree is directed and acyclic, $\bA$ is an upper triangular matrix with diagonal elements zero when the $X$'s are properly ordered. 
\begin{definition}
  Consider a tree, $\mathcal{T}=\{\bX,\bA\}$, defined by~\eqref{eq:model_X}, where $\bX=(X_{1},\dots,X_{p})^\top\in\mathbb{R}^{p}$ are the nodes in the tree and $\bA=(a_{ij})\in\mathbb{R}^{p\times p}$ is the adjacency matrix of $\bX$. We say that $X_{i}$ is a \textbf{parent} of $X_{j}$ if $a_{ij}\neq 0$, and, naturally, if $X_{i}$ is a parent of $X_{j}$, $X_{j}$ is a \textbf{child} of $X_{i}$, for $i,j=1,\dots,p$ and $i\neq j$. Let $\mathcal{P}_{j}=\{i:a_{ij}\neq 0, i=1,\dots,p\}$ denote the set of parent nodes of $X_{j}$ and $\mathcal{C}_{j}=\{k:a_{jk}\neq 0, k=1,\dots,p\}$ denote the set of child nodes of $X_{j}$, for $j=1,\dots,p$. If $\mathcal{P}_{j}=\emptyset$, $X_{j}$ is called a \textbf{root node}. If $\mathcal{C}_{j}=\emptyset$, $X_{j}$ is called a \textbf{leaf node} (or terminal node). If a node has both parent and child nodes, that is, $\mathcal{P}_{j}\neq\emptyset$ and $\mathcal{C}_{j}\neq\emptyset$, then it is called an \textbf{internal node}.
\end{definition}
Figure~\ref{fig:method_eg1} shows the tree structure of a toy example with $p=6$ predictors. In this example, $X_{1}$ is a root node, $X_{4}$, $X_{5}$ and $X_{6}$ are leaf nodes, and $X_{2}$ and $X_{3}$ are internal nodes. From Model~\eqref{eq:model_X}, 
\begin{equation}\label{eq:model_Xe}
  \bX=\bvarepsilon(\boldsymbol{\mathrm{I}}-\bA)^{-1}.
\end{equation}
Let $\bD=(\boldsymbol{\mathrm{I}}-\bA)^{-1}=(d_{ij})\in\mathbb{R}^{p\times p}$, which is also an upper triangular matrix, but with diagonal elements equal to one. The matrix $\bD$ is called the \textit{influence matrix}~\citep{shojaie2009analysis}, where the value of $d_{ij}$ quantifies the overall influence of $X_{i}$ on $X_{j}$ if $X_{i}$ is a parent of $X_{j}$. Under~\eqref{eq:model_Xe}, the covariance matrix of $\bx_{i}$ is $\bD^\top\bOmega\bD$ (for $i=1,\dots,n$). As $\bOmega$ is a diagonal matrix, this demonstrates that the dependencies among the $p$ predictors are fully characterized through $\bD$.
In the motivated sMRI example, the hierarchical brain segmentation in Figure~\ref{fig:brain_parcel} has two trees, where the two Level-0 regions are the root nodes. The direction is from Level 0 to Level 5 as the child regions are a segmentation of the parent region with potential measurement errors.

In Model~\eqref{eq:model}, $\beta_{j}$ is interpreted as the effect of $X_{j}$ on $Y$ conditional on the rest predictors, or the \emph{direct effect} $X_{j}$ on $Y$. 
\begin{proposition} 
  Under Model~\eqref{eq:model}, for $j=1,\dots,p$,
  \[
    \beta_{j}=0 \quad \Leftrightarrow\quad X_{j}~\indep~Y~|~\{X_{1},\dots,X_{j-1},X_{j+1},\dots,X_{p}\}.
  \]
\end{proposition}
\noindent With a tree structure, the \emph{total effect} of a node can also be of great interest, where the total effect accounts for all the possible path effects to the outcome. For example, in the ADNI application, for a root/internal brain region at higher levels, the existence of a total effect may suggest a global impact of atrophy on the declination of memory. Otherwise, if only the leaf nodes have a total/direct effect, the impact is localized. Therefore, we introduce the following regularization criterion to estimate the parameters.
\begin{equation}\label{eq:beta_solution}
  \hat{\bbeta}=\underset{\bbeta\in\mathbb{R}^{p}}{\arg\min}~\frac{1}{2}\|\bY-\bX\bbeta\|_{2}^{2}+\lambda\left(\|\bD\bbeta\|_{1}+\alpha\|\bbeta\|_{1}\right),
\end{equation}
where $\lambda,\alpha\geq 0\in\mathbb{R}$ are the tuning parameters. Denote
\begin{equation}
  \mathcal{R}_{1}(\bbeta;\bD)=\|\bD\bbeta\|_{1}, ~ \mathcal{R}_{2}(\bbeta)=\|\bbeta\|_{1}, \text{ and } \mathcal{R}(\bbeta;\bD)=\mathcal{R}_{1}(\bbeta;\bD)+\alpha\mathcal{R}_{2}(\bbeta).
\end{equation}
$\mathcal{R}_{1}$ regulates the total effect of the predictors, $\mathcal{R}_{2}$ regulates the direct effect, and $\mathcal{R}$ leverages both with a tuning parameter $\alpha$ . For the toy example of Figure~\ref{fig:method_eg1},
\begin{eqnarray*}
  \mathcal{R}_{1}(\bbeta;\bD) &=& |\beta_{1}+a_{12}\beta_{2}+a_{13}\beta_{3}+a_{12}a_{24}\beta_{4}+a_{12}a_{25}\beta_{5}+a_{13}a_{36}\beta_{6}| \\
  && +|\beta_{2}+a_{24}\beta_{4}+a_{25}\beta_{5}|+|\beta_{3}+a_{36}\beta_{6}| \\
  && +|\beta_{4}|+|\beta_{5}|+|\beta_{6}|,
\end{eqnarray*}
where the first line is the total effect of the root node ($X_{1}$), the second line is the total effect of the internal nodes ($X_{2}$ and $X_{3}$), and the third line is the total effect of the leaf nodes ($X_{4}$, $X_{5}$ and $X_{6}$). For a root or internal node, the total effect counts all the possible path effects to the outcome. The solution under $\mathcal{R}_{1}$ can be solved by the so-called \emph{generalized lasso}~\citep{tibshirani2011solution} by replacing the penalty matrix with the influence matrix. For $\alpha\neq 0$, let
\begin{equation}
  \tilde{\bD}=\begin{pmatrix}
    \bD \\
    \alpha\boldsymbol{\mathrm{I}}_{p}
  \end{pmatrix}\in\mathbb{R}^{2p\times p},
\end{equation}
where $\boldsymbol{\mathrm{I}}_{p}$ is the $p$-dimensional identity matrix. Then $\mathcal{R}(\bbeta;\bD)=\|\tilde{\bD}\bbeta\|_{1}$, which can also be solved by the generalized lasso.
In practice, $\bD$ can be predefined based on domain knowledge or estimated from the data. As the tree structure is defined, by properly ordering the columns of $\bX$, $\bD$ can be estimated by obtaining the covariance or precision matrix of $\bX$ first followed by a Cholesky decomposition~\citep{shojaie2010penalized}.

\begin{figure}
  \begin{center}
    \subfloat[$\mathcal{R}_{1}(\bbeta;\bD)$]{\includegraphics[width=0.3\textwidth]{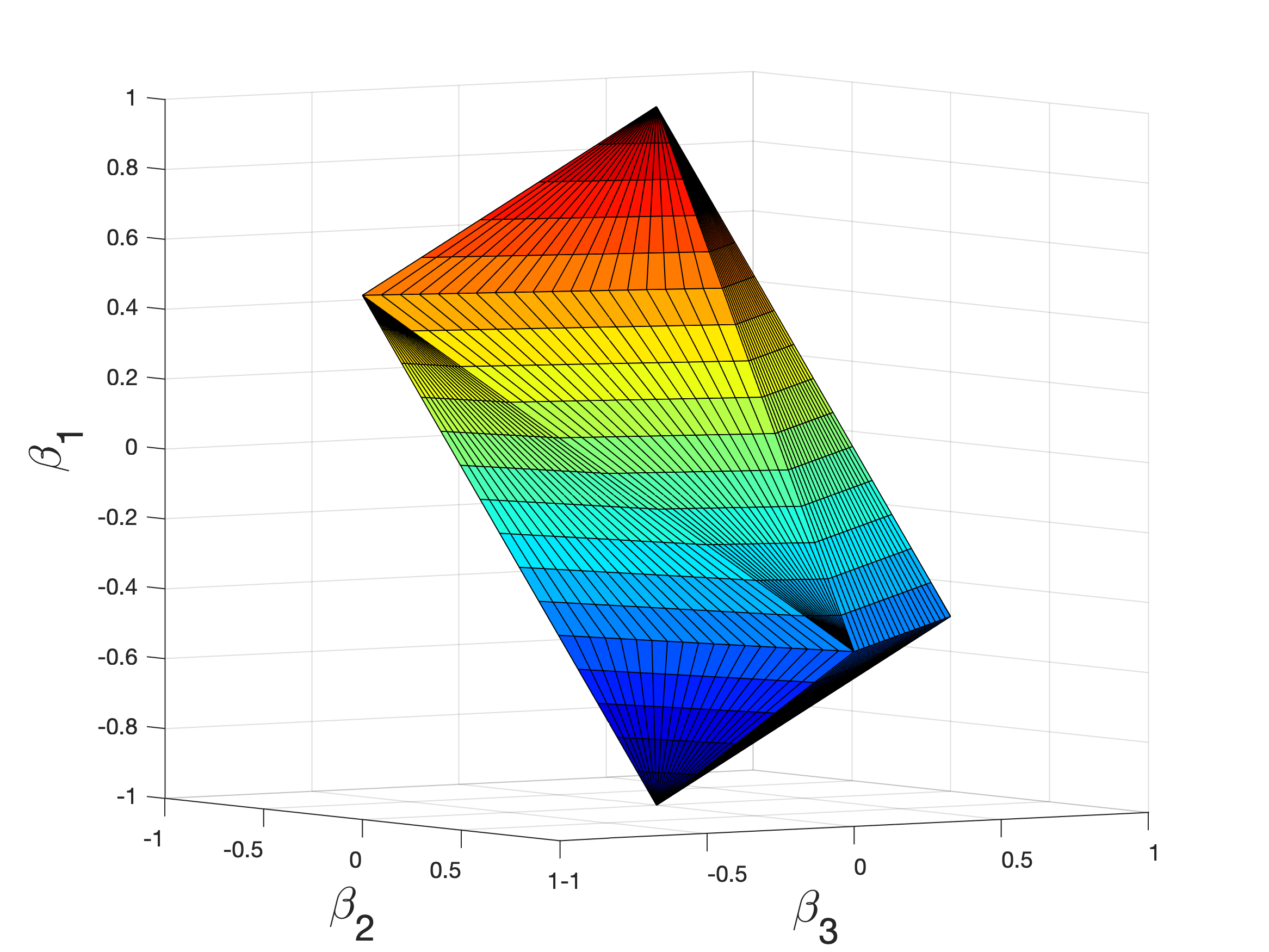}}
    \enskip{}
    \subfloat[$\mathcal{R}_{2}(\bbeta)$]{\includegraphics[width=0.3\textwidth]{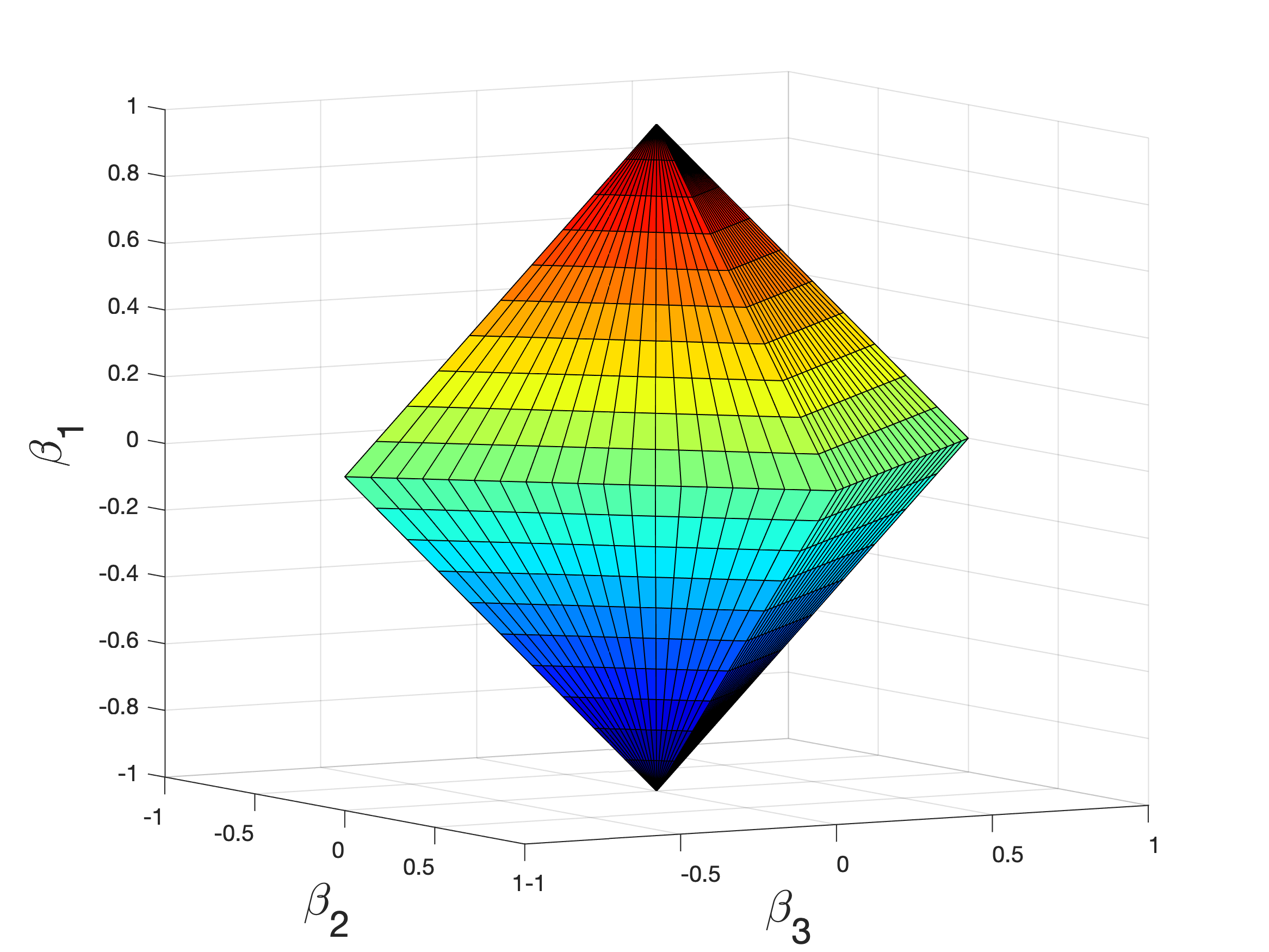}}
    \enskip{}
    \subfloat[$\mathcal{R}_{1}(\bbeta;\bD)+\mathcal{R}_{2}(\bbeta)$]{\includegraphics[width=0.3\textwidth]{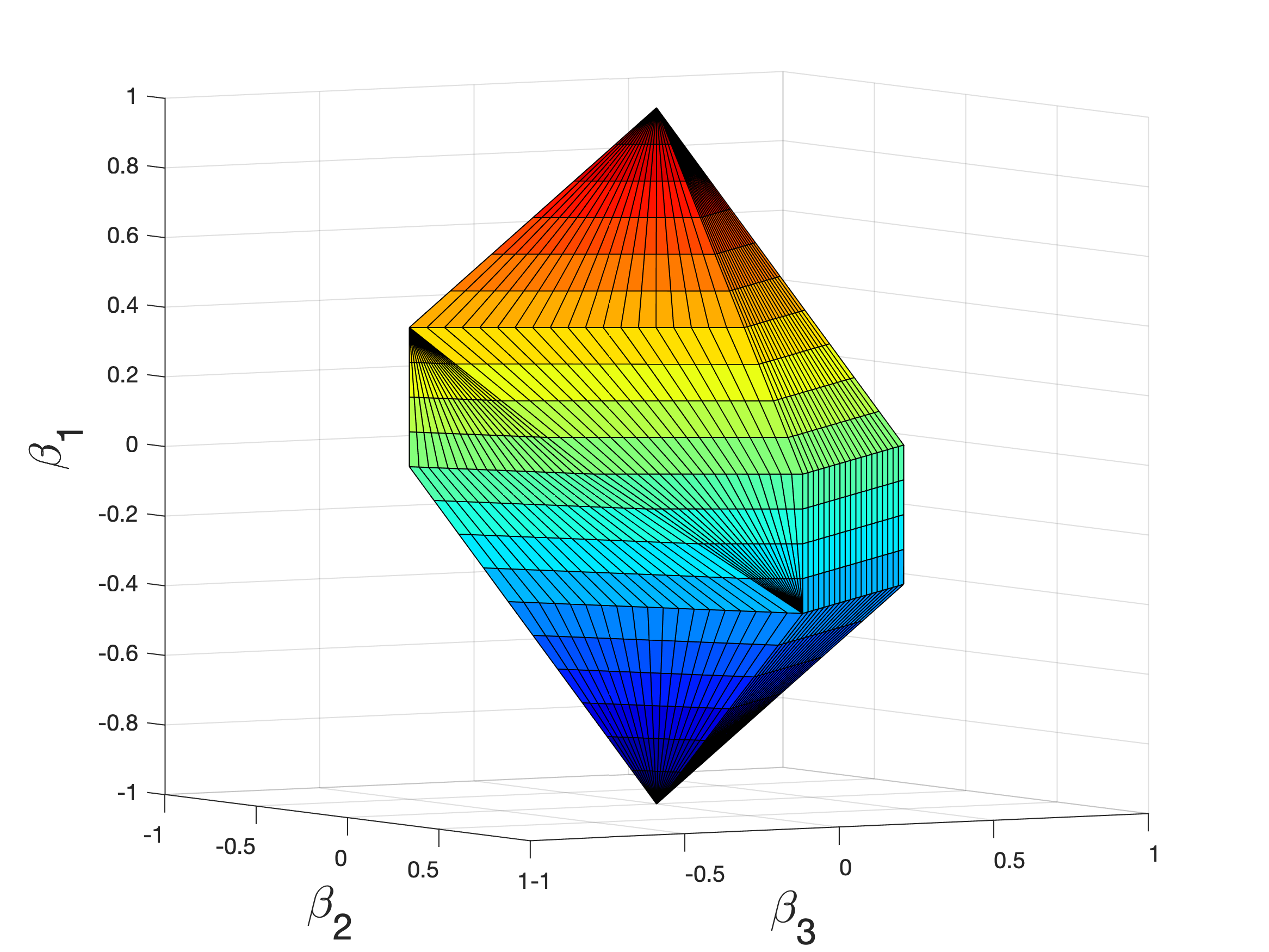}}
  \end{center}
  \caption{\label{fig:method_Rcontour}Contour plot of the regularity functions (a) $\mathcal{R}_{1}$, (b) $\mathcal{R}_{2}$, and (c) $\mathcal{R}$ (with $\alpha=1$) for the example in Figure~\ref{fig:method_eg1} with $a_{12}=a_{13}=0.5$.}
\end{figure}

For visualization purposes, we consider a simplified example with $p=3$ nodes, as shown in Figure~\ref{fig:method_eg2} and demonstrate the difference between the regularity functions. Figure~\ref{fig:method_Rcontour} shows the contour plots. Under a lasso regularization ($\mathcal{R}_{2}$), the three predictors are penalized in an equivalent way. Under $\mathcal{R}_{1}$ and $\mathcal{R}$, $X_{1}$ is regularized differently from $X_{2}$ and $X_{3}$, given the fact that $X_{1}$ is the parent of $X_{2}$ and $X_{3}$. A major drawback of the original lasso is that when the covariates are dependent, model selection is not consistent~\citep{zou2005regularization}. Taking the tree structure into consideration, $\mathcal{R}_{1}$ (or $\mathcal{R}$) circumvents this issue. This can be shown by combining models~\eqref{eq:model} and~\eqref{eq:model_Xe}, 
\begin{equation}
  \bY=\bX\bbeta+\bepsilon=\bvarepsilon\bD\bbeta+\bepsilon\triangleq\bvarepsilon\bgamma+\bepsilon,
\end{equation}
where $\bgamma=\bD\bbeta\in\mathbb{R}^{p}$ is the new model parameter. Under this definition,
\begin{equation}
  \mathcal{R}_{1}(\bbeta;\bD)=\|\bD\bbeta\|_{1}=\|\bgamma\|_{1}.
\end{equation}
Solving the optimization problem~\eqref{eq:beta_solution} with $\mathcal{R}_{1}$ as the regularization becomes equivalent to searching for an ordinary lasso solution with $\bvarepsilon$ as the design matrix, where the columns of $\bvarepsilon$ are mutually independent. When $\bD$ is correctly specified, the estimated active set under $\mathcal{R}_{1}$ is consistent. In the next section, we show that when the design matrix is well behaved corresponding to the regularization matrix $\tilde{\bD}$ in $\mathcal{R}$, the estimator of $\bbeta$ is consistent in $\ell_{2}$-norm and model selection is also consistent.


\subsection{Estimation consistency and model selection consistency}
\label{sub:method_consist}

Let $\bbeta^{*}$ denote the true model parameter and  $\mathcal{M}=\{\bbeta\in\mathbb{R}^{p}~|~(\tilde{\bD}\bbeta)_{\mathcal{S}^{c}}=0\}$ denote the model space under regularization $\mathcal{R}$, where $\mathcal{S}$ is the support of $\tilde{\bD}\bbeta^{*}$ and $\mathcal{S}^{c}$ is the complement of $\mathcal{S}$. In order to achieve estimation consistency and model selection consistency, the following assumptions are imposed. The first assumption is on the sample \textit{Fisher information matrix}, $Q=\nabla^{2}\ell(\bbeta^{*})$, where $\ell$ is the loss function and $\nabla$ is the differential operator. Under~\eqref{eq:beta_solution}, $\ell=\|\bY-\bX\bbeta\|_{2}^{2}/2$. The second is also on $Q$, but with respect to the regularization matrix $\tilde{\bD}$.
\begin{description}
  \item[Assumption 1] (Restricted strong convexity, RSC) Let $\mathcal{C}\subset\mathbb{R}^{p}$ be a known convex set containing $\bbeta^{*}$. The loss function $\ell$ is RSC on $\mathcal{C}\cap\mathcal{M}$ when
  \begin{equation*}
    \btheta^\top \nabla^{2}\ell(\bbeta)\btheta\geq m\|\btheta\|_{2}^{2}, ~\bbeta\in\mathcal{C}\cap\mathcal{M}, ~\btheta\in(\mathcal{C}\cap\mathcal{M})-(\mathcal{C}\cap\mathcal{M}),
  \end{equation*}
  \begin{equation*}
    \|\nabla^{2}\ell(\bbeta)-Q\|_{2}\leq L\|\bbeta-\bbeta^{*}\|_{2}, ~\bbeta\in\mathcal{C},
  \end{equation*}
  for some $m>0$ and $L<\infty$.
  \item[Assumption 2] For $\tau\in(0,1)$,
  \begin{equation*}
    \|\tilde{\bD}_{\mathcal{S}^{c}}\bX^\top(\tilde{\bD}_{\mathcal{S}}\bX^\top)^{-}\mathrm{sign}\{(\tilde{\bD}\bbeta^{*})_{\mathcal{S}}\}\|_{\infty}\leq 1-\tau,
  \end{equation*}
  where $\tilde{\bD}_{\mathcal{S}}\in\mathbb{R}^{|\mathcal{S}|\times p}$ takes the rows of $\tilde{\bD}$ in $\mathcal{S}$, $(\tilde{\bD}\bbeta^{*})\in\mathbb{R}^{|\mathcal{S}|}$, and $\bA^{-}$ is the \textit{Moore-Penrose pseudoinverse} of a matrix $\bA\in\mathbb{R}^{p\times p}$ and $\mathrm{sign}(\cdot)$ is the sign function.
\end{description}
For a sparse regression problem like~\eqref{eq:beta_solution}, with random Gaussian or sub-Gaussian designs, the RSC condition is satisfied, even when the predictors are dependent~\citep{raskutti2010restricted,rudelson2012reconstruction}. Assumption 2 is an irrepresentability condition that requires the active predictors (with respect to $\tilde{\bD}$) to be not overly well-aligned with the inactive predictors. The ideal scenario is that the inactive predictors are orthogonal to the active predictors, which is impossible to realize when the data are high-dimensional. Assumption 2 relaxes the orthogonality to near orthogonality.
The following theorem is an adaption of Corollary 4.2 in \citet{lee2015model} to the considered regularization $\mathcal{R}$ when $\alpha\neq0$.
\begin{theorem}\label{thm:consist}
  Assume $\alpha\neq0$. Under Assumptions 1 and 2, for some $0<\kappa_{1},\kappa_{2},\kappa_{3}<\infty$ and $\lambda=(8\kappa_{1}\sigma/\tau)\sqrt{\log{p}/n}$, the estimator under $\mathcal{R}$ is unique, and with probability $1-2p^{-1}$,
  \begin{enumerate}[(1)]
    \item consistent:
      \[
        \|\hat{\bbeta}-\bbeta^{*}\|_{2}\leq\frac{4}{m}\left(\kappa_{3}+4\kappa_{1}\kappa_{2}/\tau\right)\sigma\sqrt{\frac{\log{p}}{n}},
      \]
    \item model selection consistent:
      \[
        \hat{\bbeta}\in\mathcal{M}.
      \]
  \end{enumerate}
\end{theorem}
The proof of Theorem~\ref{thm:consist} and values of $\kappa_{1},\kappa_{2},\kappa_{3}$ are provided in Appendix Section~\ref{appendix:sub:proof_thm_consist}. When $\alpha=0$, an analogous consistency holds, where the compatibility constants, $\kappa_{1},\kappa_{2},\kappa_{3}$, are computed with respect to the regularization function $\mathcal{R}_{1}$.

\subsection{Algorithm}
\label{sub:method_alg}

As discussed above, the proposed regularizations, $\mathcal{R}_{1}$ and $\mathcal{R}$, can be solved through the generalized lasso~\citep{tibshirani2011solution}. The ordinary lasso, $\mathcal{R}_{2}$, can also be solved by the generalized lasso by setting the sparsity matrix to be the $p$-dimensional identity matrix. Thus, the algorithm introduced in \citet{tibshirani2011solution} will be employed to estimate the model parameters, though in the generalized lasso literature, the proposed tree formulation was not considered. To choose the tuning parameters, the $C_{p}$ criterion is considered, which is defined as
\begin{equation}
  C_{p}(\lambda)=\|\bY-\bX\hat{\bbeta}_{\lambda}\|_{2}^{2}-n\sigma^{2}+2\sigma^{2}\mathrm{df}(\bX\hat{\bbeta}_{\lambda}),
\end{equation}
where $\hat{\bbeta}_{\lambda}$ is the estimate of $\bbeta$ under parameter $\lambda$ and $\mathrm{df}$ is the degrees of freedom. An unbiased estimate of $C_{p}$ is provided by \citet{tibshirani2011solution}. It is suggested to choose the $\lambda$ that minimizes $\hat{C}_{p}(\lambda)$.

\section{Simulation Study}
\label{sec:sim}

\begin{figure}
  \begin{center}
    \subfloat[]{\includegraphics[width=0.23\textwidth,page=3]{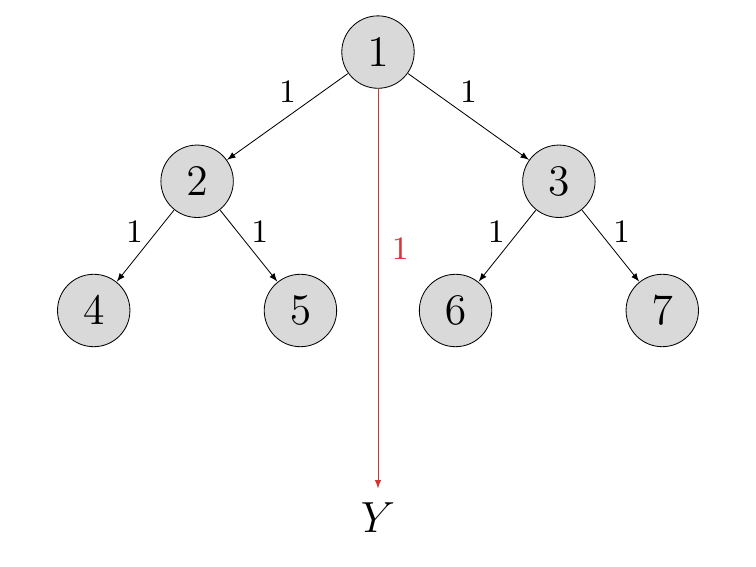}}
    \enskip{}
    \subfloat[]{\includegraphics[width=0.23\textwidth,page=4]{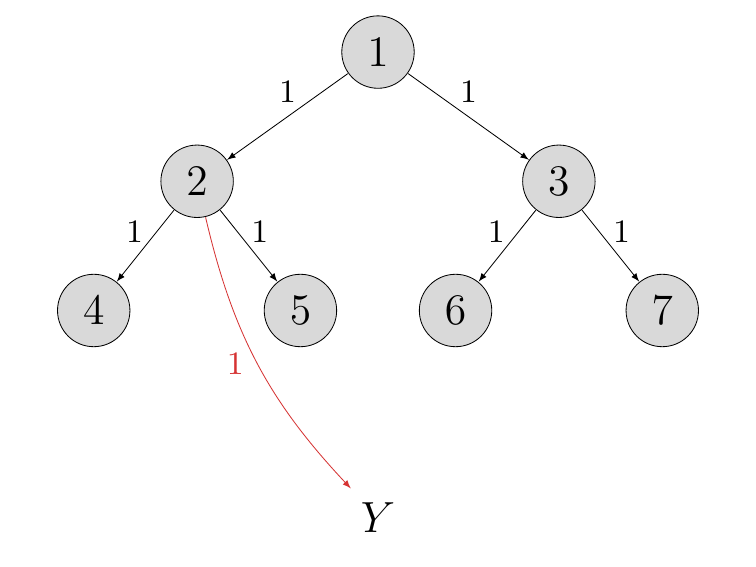}}
    \enskip{}
    \subfloat[]{\includegraphics[width=0.23\textwidth,page=5]{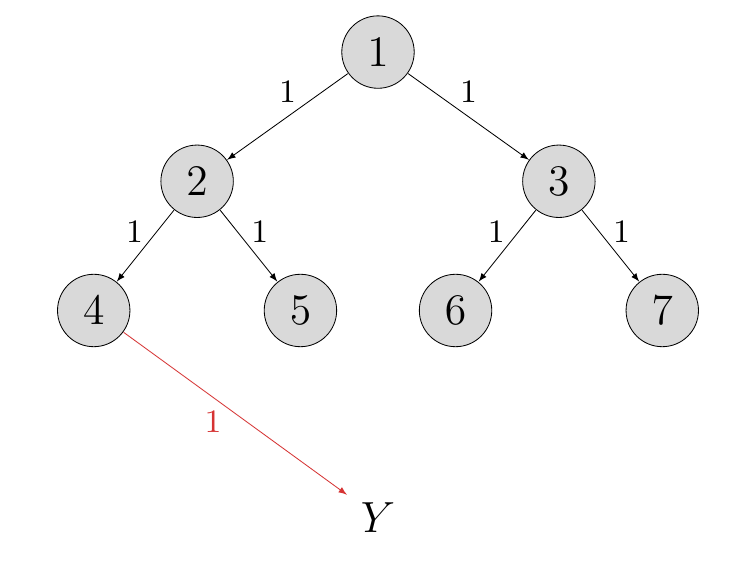}}
    \enskip{}
    \subfloat[]{\includegraphics[width=0.23\textwidth,page=6]{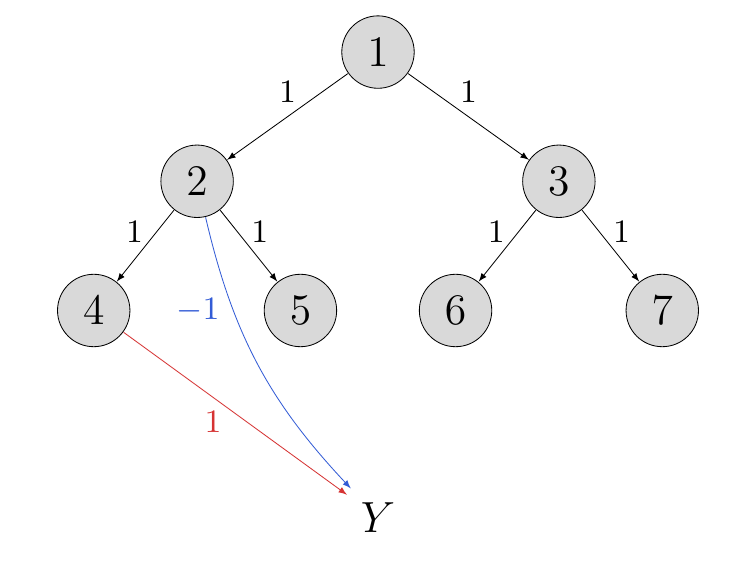}}
  \end{center}
  \caption{\label{fig:sim_p7_model}The considered model specification in the simulation study when $p=7$.}
\end{figure}

\begin{figure}
  \begin{center}
    \subfloat[]{\includegraphics[width=0.3\textwidth,page=7]{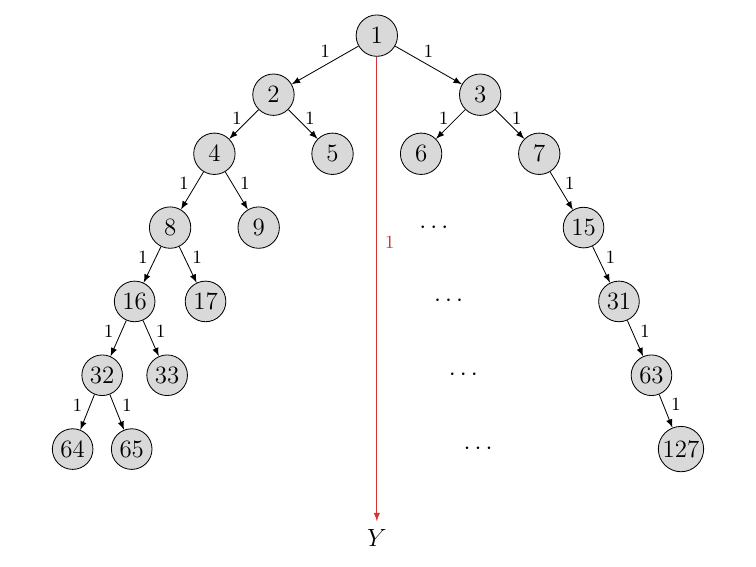}}
    \enskip{}
    \subfloat[]{\includegraphics[width=0.3\textwidth,page=8]{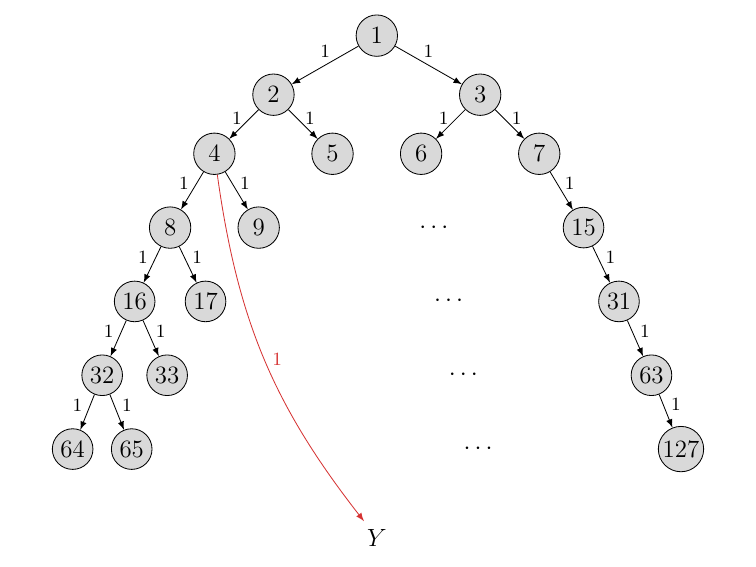}}
    \enskip{}
    \subfloat[]{\includegraphics[width=0.3\textwidth,page=9]{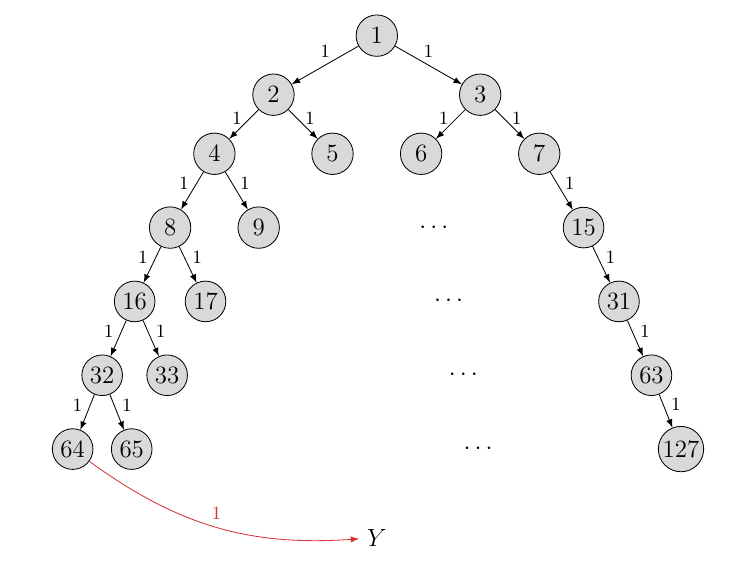}}
    
    \subfloat[]{\includegraphics[width=0.3\textwidth,page=10]{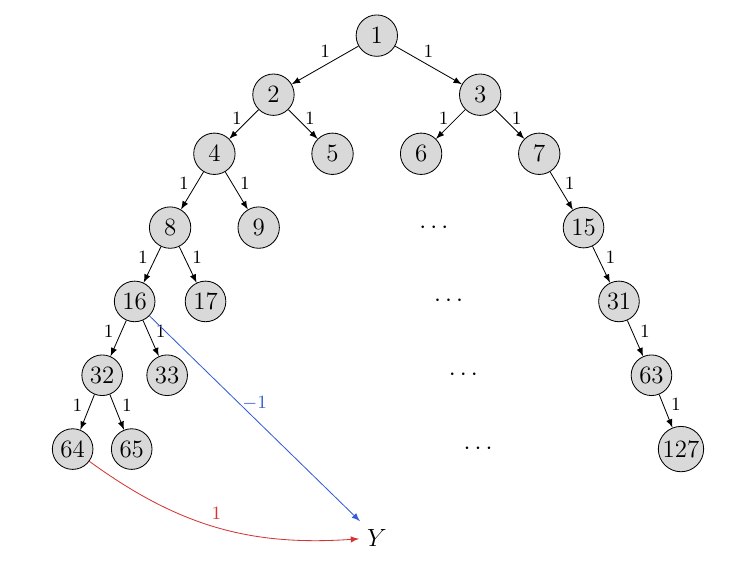}}
    \enskip{}
    \subfloat[]{\includegraphics[width=0.3\textwidth,page=11]{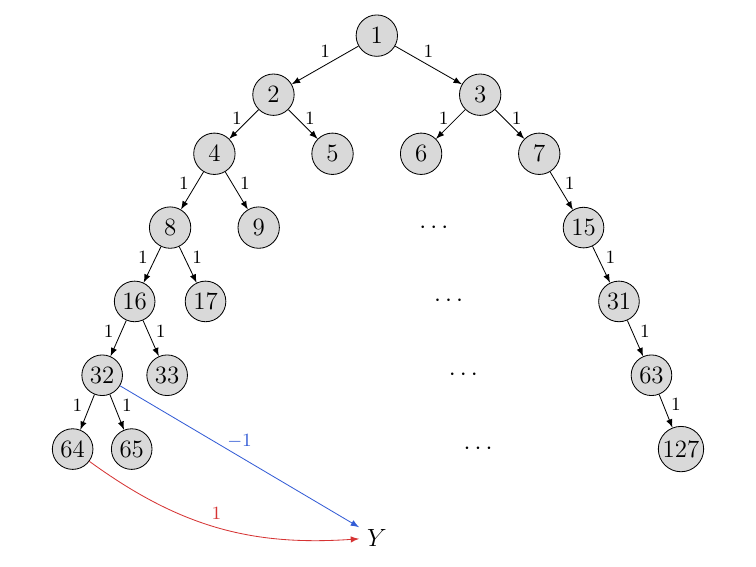}}
    \enskip{}
    \subfloat[]{\includegraphics[width=0.3\textwidth,page=12]{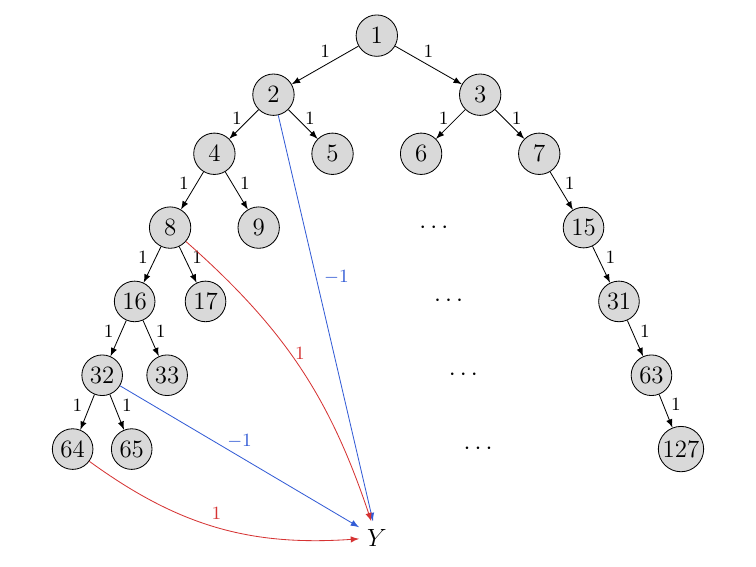}}
  \end{center}
  \caption{\label{fig:sim_p127_model}The considered model specification in the simulation study when $p=127$.}
\end{figure}

In the simulation study, we consider binary trees, where each root or internal node has two children. Two cases are considered, (1) $L=3$ levels with $p=7$ nodes and (2) $L=7$ levels with $p=127$ nodes. $X$'s are first generated following~\eqref{eq:model_X}, where the nonzero elements of the adjacency matrix are set to be one. For $\bbeta$, various scenarios are considered. Figures~\ref{fig:sim_p7_model} and~\ref{fig:sim_p127_model} present the scenarios when $p=7$ and $p=127$, respectively. All errors in~\eqref{eq:model} and~\eqref{eq:model_X} are independently generated from the standard normal distribution. Four types of regularization are considered: (i) $\mathcal{R}_{1}$, (ii) $\mathcal{R}_{1}+\alpha\mathcal{R}_{2}$, (iii) $\mathcal{R}_{2}$ (the Lasso regularization), and (iv) the elastic net \cite[EN,][]{zou2005regularization}. For (ii), various values of $\alpha$ are considered; and for (iv), multiple choices of the $\ell_{2}$ proportion are considered. For both cases, we present the results under the value chosen by $C_{p}$ introduced in Section~\ref{sub:method_alg}. For all the approaches, $\lambda$ is a tuning parameter chosen by $C_{p}$. A sample size of $n=50$ is considered and the simulation is repeated for 200 replications. In the implementation, the value of $\bD$ is calculated from the adjacency matrix and imputed into the regularization function. The estimate of $\bbeta$ is obtained directly from the approaches. The estimate of $\bgamma$ is also obtained using the definition, $\hat{\bgamma}=\bD\hat{\bbeta}$, where $\hat{\bbeta}$ is the estimate of $\bbeta$ and $\hat{\bgamma}$ is the estimate of $\bgamma$. To evaluate the performance, the sensitivity and specificity of identifying nonzero parameters are considered, as well as the mean squared error (MSE), defined as
\begin{equation}
  \mathrm{MSE}(\hat{\bbeta})=\mathbb{E}\|\hat{\bbeta}-\bbeta\|_{2}^{2}=\mathbb{E}\left\{\sum_{j=1}^{p}(\hat{\beta}_{j}-\beta_{j})^{2}\right\}.
\end{equation}
The MSE of $\bgamma$ estimate is defined analogously. An estimate of the MSE is acquired by averaging over the 200 replications.

In Tables~\ref{table:sim_p7} and~\ref{table:sim_p127}, we present the results with $p=7$ and $p=127$, respectively. From the tables, when only the leaf node has a direct effect on the outcome (case (c) in Tables~\ref{table:sim_p7} and~\ref{table:sim_p127}), the lasso ($\mathcal{R}_{2}$) and the EN regularization perform better in estimating $\bbeta$ and $\bgamma$ with higher sensitivity and specificity and lower MSE. When the root or internal node has a direct effect on the outcome (cases (a) and (b) in Tables~\ref{table:sim_p7} and~\ref{table:sim_p127}), the EN performs slightly better in estimating $\bbeta$, while $\mathcal{R}_{1}$ performs slightly better when estimating $\bgamma$. This is anticipated, as $\mathcal{R}_{1}$ aims to regularize the total effect. Case (d) in Table~\ref{table:sim_p7} and cases (d)--(f) in Table~\ref{table:sim_p127} consider scenarios where an internal node has a nonzero direct effect on the outcome, while the total effect is zero. Under this scenario, the specificity of $\mathcal{R}_{2}$ and EN are dramatically low, especially in estimating $\bgamma$. For the high-dimensional case when $p=127$, the sensitivity of $\mathcal{R}_{2}$ in identifying nonzero $\bbeta$ becomes much lower than the other two approaches, due to the high collinearity between the predictors. By defining the regularization based on the tree structure, the proposed regularizations, $\mathcal{R}_{1}$ and $\mathcal{R}$, circumvent this issue. In general, $\mathcal{R}_{1}$ yields better performance in estimating $\bgamma$. $\mathcal{R}$ leverages the advantage of $\mathcal{R}_{2}$ and slightly improves the performance for estimating $\bbeta$.


\begin{table}
  \caption{\label{table:sim_p7}The average estimated $C_{p}$, sensitivity and specificity, and the mean squared error (MSE) of $\bbeta$ and $\bgamma$ estimation in the simulation study with $p=7$ and $n=50$. The results are the averages over 200 replications.}
  \begin{center}
    \begin{tabular}{c l c c r c c c r}
      \hline
      & & \multicolumn{3}{c}{$\beta$} && \multicolumn{3}{c}{$\gamma$} \\
      \cline{3-5}\cline{7-9}
      \multirow{-2}{*}{Model} & \multicolumn{1}{c}{\multirow{-2}{*}{Method}} & \multicolumn{1}{c}{Sensitivity} & \multicolumn{1}{c}{Specificity} & \multicolumn{1}{c}{MSE} && \multicolumn{1}{c}{Sensitivity} & \multicolumn{1}{c}{Specificity} & \multicolumn{1}{c}{MSE} \\
      \hline
      & $\mathcal{R}_{1}$ & 1.000 & 0.669 & 0.113 && 1.000 & 0.755 & 0.075 \\
      & $\mathcal{R}$ & 0.980 & 0.698 & 0.162 && 0.965 & 0.661 & 0.115 \\
      & Lasso ($\mathcal{R}_{2}$) & 1.000 & 0.575 & 0.170 && 1.000 & 0.423 & 0.101 \\
      \multirow{-4}{*}{(a)} & EN & 1.000 & 0.741 & 0.121 && 1.000 & 0.649 & 0.076 \\
      \hline
      & $\mathcal{R}_{1}$ & 1.000 & 0.536 & 0.162 && 1.000 & 0.711 & 0.108 \\
      & $\mathcal{R}$ & 0.995 & 0.712 & 0.165 && 0.985 & 0.707 & 0.119 \\
      & Lasso ($\mathcal{R}_{2}$) & 1.000 & 0.658 & 0.140 && 1.000 & 0.550 & 0.092 \\
      \multirow{-4}{*}{(b)} & EN & 1.000 & 0.726 & 0.136 && 1.000 & 0.693 & 0.090 \\
      \hline
      & $\mathcal{R}_{1}$ & 1.000 & 0.363 & 0.200 && 1.000 & 0.646 & 0.130 \\
      & $\mathcal{R}$ & 0.970 & 0.684 & 0.220 && 0.977 & 0.699 & 0.168 \\
      & Lasso ($\mathcal{R}_{2}$) & 1.000 & 0.797 & 0.068 && 1.000 & 0.702 & 0.069 \\
      \multirow{-4}{*}{(c)} & EN & 1.000 & 0.752 & 0.083 && 1.000 & 0.729 & 0.077 \\
      \hline
      & $\mathcal{R}_{1}$ & 1.000 & 0.658 & 0.145 && 1.000 & 0.762 & 0.076 \\
      & $\mathcal{R}$ & 0.980 & 0.652 & 0.213 && 0.970 & 0.724 & 0.108 \\
      & Lasso ($\mathcal{R}_{2}$) & 0.997 & 0.522 & 0.223 && 1.000 & 0.261 & 0.125 \\
      \multirow{-4}{*}{(d)} & EN & 1.000 & 0.459 & 0.253 && 1.000 & 0.252 & 0.136 \\
      \hline
    \end{tabular}
  \end{center}
\end{table}
\begin{table}
  \caption{\label{table:sim_p127}The average estimated $C_{p}$, sensitivity, specificity, as well as the estimation mean squared error (MSE) in the simulation study with $p=127$ and $n=50$. The results are the averages over 200 replications.}
  \begin{center}
    \begin{tabular}{c l c c r c c c r}
      \hline
      & & \multicolumn{3}{c}{$\beta$} && \multicolumn{3}{c}{$\gamma$} \\
      \cline{3-5}\cline{7-9}
      \multirow{-2}{*}{Model} & \multicolumn{1}{c}{\multirow{-2}{*}{Method}} & \multicolumn{1}{c}{Sensitivity} & \multicolumn{1}{c}{Specificity} & \multicolumn{1}{c}{MSE} && \multicolumn{1}{c}{Sensitivity} & \multicolumn{1}{c}{Specificity} & \multicolumn{1}{c}{MSE} \\
      \hline
      & $\mathcal{R}_{1}$ & 1.000 & 0.901 & 0.261 && 1.000 & 0.947 & 0.181 \\
      & $\mathcal{R}$ & 0.965 & 0.919 & 0.245 && 0.965 & 0.925 & 0.194 \\
      & Lasso ($\mathcal{R}_{2}$) & 0.735 & 0.894 & 0.618 && 1.000 & 0.660 & 0.558 \\
      \multirow{-4}{*}{(a)} & EN & 1.000 & 0.958 & 0.178 && 1.000 & 0.850 & 0.173 \\
      \hline
      & $\mathcal{R}_{1}$ & 1.000 & 0.818 & 0.568 && 1.000 & 0.906 & 0.497 \\
      & $\mathcal{R}$ & 0.955 & 0.902 & 0.471 && 0.948 & 0.849 & 0.515 \\
      & Lasso ($\mathcal{R}_{2}$) & 0.975 & 0.906 & 0.326 && 1.000 & 0.702 & 0.396 \\
      \multirow{-4}{*}{(b)} & EN & 0.995 & 0.950 & 0.165 && 1.000 & 0.853 & 0.201 \\
      \hline
      & $\mathcal{R}_{1}$ & 0.990 & 0.686 & 1.695 && 0.991 & 0.837 & 1.506 \\
      & $\mathcal{R}$ & 0.980 & 0.850 & 0.926 && 0.942 & 0.786 & 1.159 \\
      & Lasso ($\mathcal{R}_{2}$) & 1.000 & 0.958 & 0.055 && 1.000 & 0.851 & 0.175 \\
      \multirow{-4}{*}{(c)} & EN & 1.000 & 0.955 & 0.076 && 1.000 & 0.867 & 0.194 \\
      \hline
      & $\mathcal{R}_{1}$ & 1.000 & 0.862 & 0.500 && 1.000 & 0.929 & 0.337 \\
      & $\mathcal{R}$ & 0.970 & 0.898 & 0.442 && 0.970 & 0.908 & 0.341 \\
      & Lasso ($\mathcal{R}_{2}$) & 0.945 & 0.868 & 0.667 && 1.000 & 0.581 & 0.743 \\
      \multirow{-4}{*}{(d)} & EN & 0.945 & 0.869 & 0.769 && 1.000 & 0.615 & 0.793 \\
      \hline
      & $\mathcal{R}_{1}$ & 1.000 & 0.906 & 0.384 && 1.000 & 0.950 & 0.192 \\
      & $\mathcal{R}$ & 0.972 & 0.898 & 0.479 && 0.970 & 0.938 & 0.245 \\
      & Lasso ($\mathcal{R}_{2}$) & 0.760 & 0.853 & 1.272 && 0.945 & 0.556 & 0.998 \\
      \multirow{-4}{*}{(e)} & EN & 0.717 & 0.869 & 1.467 && 0.945 & 0.616 & 1.011 \\
      \hline
      & $\mathcal{R}_{1}$ & 1.000 & 0.814 & 0.880 && 1.000 & 0.900 & 0.517 \\
      & $\mathcal{R}$ & 0.961 & 0.814 & 1.149 && 0.955 & 0.886 & 0.683 \\
      & Lasso ($\mathcal{R}_{2}$) & 0.720 & 0.799 & 2.565 && 1.000 & 0.473 & 1.917 \\
      \multirow{-4}{*}{(f)} & EN & 0.755 & 0.838 & 2.383 && 0.995 & 0.569 & 1.507 \\
      \hline
    \end{tabular}
  \end{center}
\end{table}

\section{The Alzheimer's Disease Neuroimaging Initiative Study}
\label{sec:adni}

We apply the proposed approach to the MRI data collected by the Alzheimer's Disease Neuroimaging Initiative (ADNI, \url{adni.loni.usc.edu}).
The ADNI study was launched in 2003 as a public-private partnership, led by Principal Investigator Michael W. Weiner, MD. The primary goal of ADNI has been to test whether serial magnetic resonance imaging (MRI), positron emission tomography (PET), other biological markers, and clinical and neuropsychological assessments can be combined to measure the progression of mild cognitive impairment (MCI) and early Alzheimer's disease (AD). 

A total of $n=590$ subjects aged between 55 and 91 are included in this study (358 Male and 232 Female). These subjects are diagnosed with MCI (402 subjects) or AD (188 subjects) at recruitment based on the cognitive and behavioral evaluation batteries. The high-resolution T1-weighted images acquired at the initial screening are processed and segmented through MRICloud \cite[\url{www.MRICloud.org},][]{mori2016mricloud}, which is a publicly available web-based platform for multi-contrast imaging segmentation and quantification. The processing steps include (1) orientation adjustment and inhomogeneity correction, (2) initial segmentation of major tissues (white and gray matter, CSF), skull, and background, (3) skull stripping and histogram matching, (4) sequential affine transformations, (5) large deformation diffeomorphic mapping (LDDMM), and (6) multi-atlas labeling fusion (MALF) with PICSL adjustment~\citep{tang2013bayesian}. The brain volumetric data follow a hierarchical tree structure with six levels. Figure~\ref{fig:brain_parcel} presents the hierarchical data structure. 
From Level 0 to Level 5, there are $p=323$ regions: 1 in Level 0, 7 in Level 1, 13 in Level 2, 44 in Level 3, 108 in Level 4, and 150 in Level 5. The multi-level volumetric data satisfies a compositional property. That is, the volume of root/internal nodes is the sum of the volume of their children. Based on this property, after standardizing the data, the $(i,j)$ element in the adjacency matrix is the ratio of the standard deviation of region $j$ over region $i$. In this study, ADNI\_MEM, which is a composite score of memory, is considered as the outcome ($Y$) to study the association between brain volume composition and memory. This composite score has been validated in \citet{crane2012development}. It uses data from the ADNI neuropsychological battery following item response theory methods, including different word lists in the Rey Auditory Verbal Learning Test and the ADAS-Cog, and by Logical Memory I data missing by design. A higher ADNI\_MEM outcome indicates better performance in the tests. We take the ADNI\_MEM score acquired on the same day as the MRI scan or the first post-imaging measurement as the outcome. 
We apply the proposed approach to investigate the association between global/local brain volume and memory.

\begin{figure}
  \begin{center}
    \subfloat[\label{fig:adni_brain_de}Direct effect]{\includegraphics[width=0.68\textwidth]{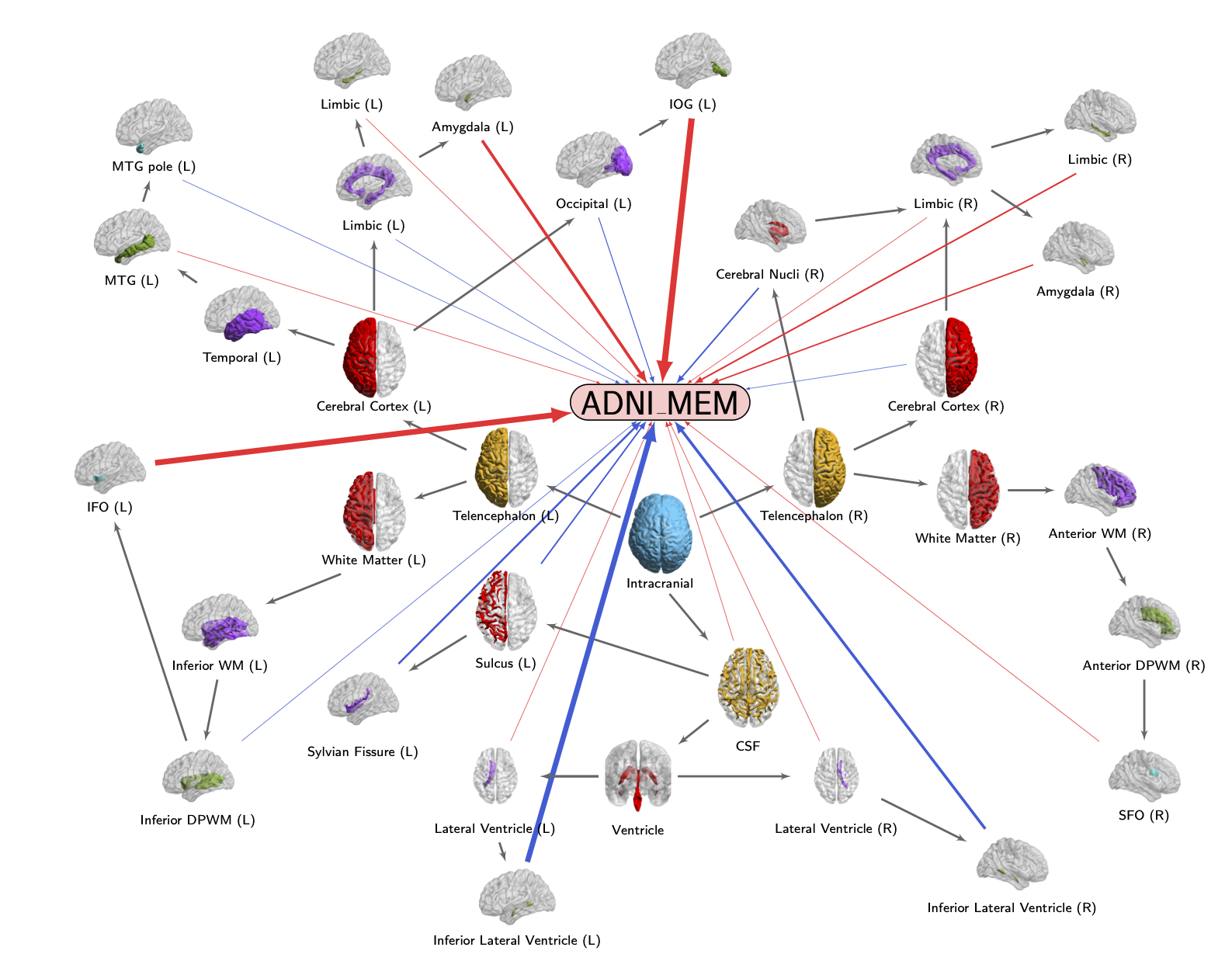}}

    \subfloat[\label{fig:adni_brain_te}Total effect]{\includegraphics[width=0.68\textwidth]{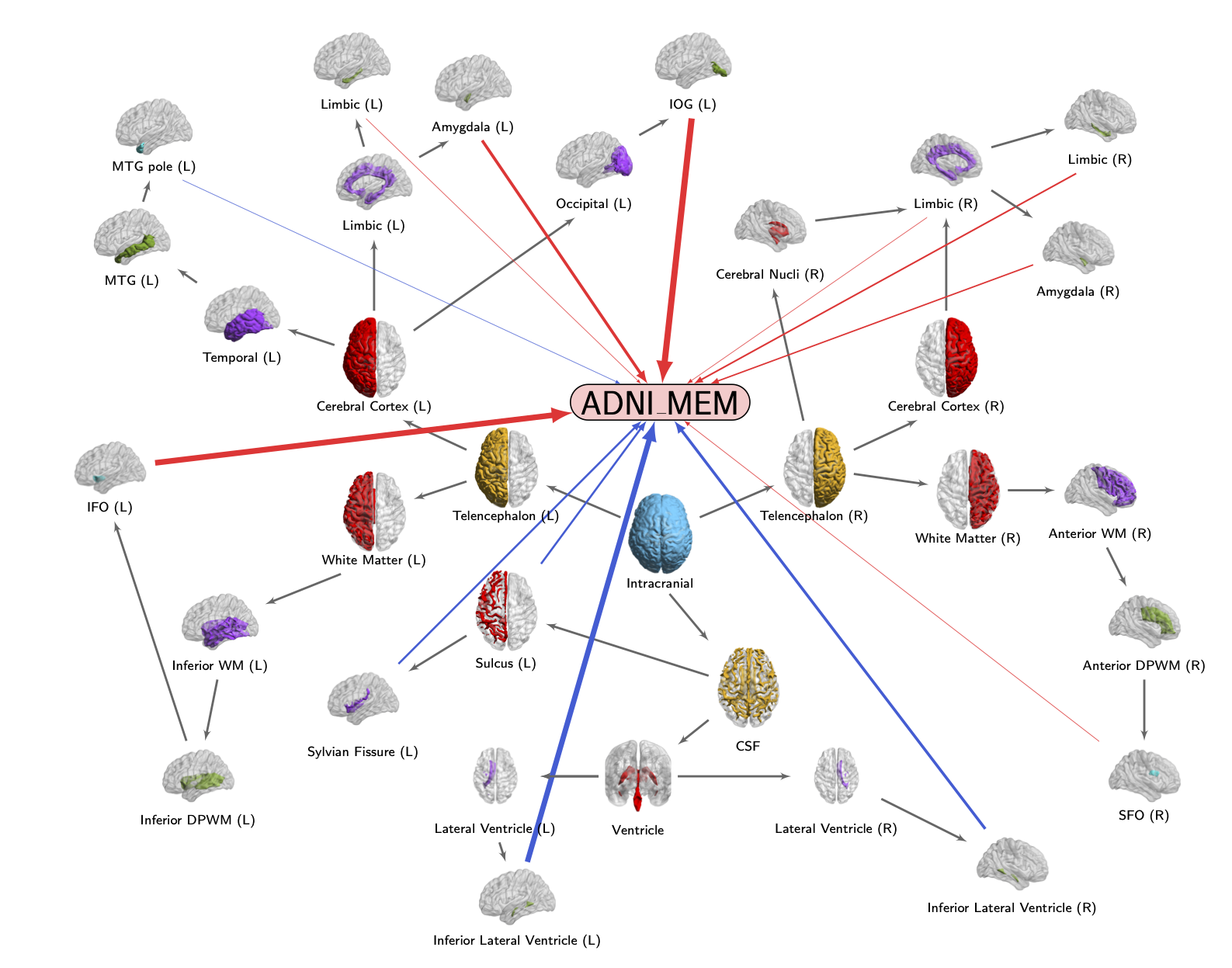}}

    \includegraphics[width=0.45\textwidth]{legend_level.png}
  \end{center}
  \vskip-25pt
  \caption{\label{fig:adni_brain}Relevant brain regions (a) with a nonzero direct effect and (b) with a nonzero total effect. Brain maps are colored by level. The gray arrows inform the hierarchical structure. A red arrow indicates a positive effect and a blue one indicates a negative effect. The width of the lines is proportional to the magnitude of the effect.}
\end{figure}

Based on the simulation results, regularization $\mathcal{R}$ is employed, where tuning parameters, $\alpha$ and $\lambda$, are chosen based on the estimated $C_{p}$. Figure~\ref{fig:adni_brain_de} presents the brain regions with a nonzero direct effect estimate and their parent brain region. In the figure, arrows in gray inform the hierarchal structure between regions. The brain maps are colored corresponding to the segmentation level. A red arrow to ADNI\_MEM indicates a positive direct effect and a blue arrow indicates a negative direct effect. Figure~\ref{fig:adni_brain_te} shows the total effect of the regions included in Figure~\ref{fig:adni_brain_de}. The estimate of the effects is presented in Table~\ref{appendix:table:adni_est} in Appendix Section~\ref{appendix:sub:adni_est}.
Compared to the results from the lasso ($\mathcal{R}_{2}$) and the elastic net (Figure~\ref{appendix:fig:adni_comp_tree} in Appendix Section~\ref{appendix:sub:adni_comp}), a more clear hierarchical structure is observed under $\mathcal{R}$. Most of the regions identified by the lasso and the elastic net are Level-4 and Level-5 regions, where the effect of some regions can be aggregated into the parent region in a higher level.

Regional brain atrophy is observed in normal aging and AD, which has been found to be associated with cognitive impairments, such as memory declination. A stereotypical pattern of neurodegeneration suggests that the atrophy occurs early in the medial temporal lobe and soon after spreads to the rest of the cortical areas following a trajectory of temporal--parietal--frontal, while motor areas are not generally impacted until the later stages of the disease~\citep{pini2016brain}. From Figure~\ref{fig:adni_brain_de}, nonzero direct effects are observed in the limbic system \cite[the part of the cerebral cortex that is beneath the temporal lobe and is involved in multiple complex functions, particularly in emotional and behavioral responses,][]{purves2004neuroscience}, the temporal and occipital lobes, and the lateral ventricles. 
Figure~\ref{fig:adni_brain_te} presents the estimated total effect using the prespecified adjacency matrix. From Figure~\ref{fig:adni_brain_te}, nonzero total effects are mainly observed in Level 4 and Level 5 regions suggesting localized impacts of atrophy on memory.

Based on the segmentation, the Level 4 limbic area consists of the parahippocampal and entorhinal cortices in Level 5. Together with the amygdala, these are all key AD markers, repeatedly verified in the existing literature~\citep{de2004mri,st2006comprehensive,jones2006differential,barnes2006measurements}. Positive direct and total effects are estimated using the proposed approach suggesting the association between atrophy in these areas and memory decline. The gray matter loss in the lateral temporal cortex, dorsal parietal and frontal cortex occurs during the progression from incipient to mild AD. During this period, cognitive deficits are observed in both memory and non-memory domains including language, visuo-spatial and executive function~\citep{frisoni2009vivo}. For the sensorimotor and visual cortices, atrophy is observed until later stages of AD~\citep{pini2016brain}. 
The left anterior insula/frontal operculum complex (IFO) also has a strong positive direct/total effect on the memory outcome. The association between atrophy in the insular cortex and cognitive deficits, such as memory, in AD has been reported in existing literature~\citep{foundas1997atrophy,lin2017insula}. 
The proposed approach identifies a positive direct/total effect of the right superior fronto-occipital fasciculus (SFO) on memory (Figure~\ref{fig:adni_brain} shows the location of the core of the SFO). In the human brain, the SFO can either be an isolated fasciculus or a branch of the superior longitudinal fasciculus~\citep{bao2017superior}. It plays a major role in speech and language, as well as the top-down modulation of visual processing and spatial aspects of cognitive processing~\citep{bar2006top,schmahmann2007association}, and was found to be associated with cognitive decline in the aging population~\citep{price2020novel}.
Negative direct/total effects are mainly observed in the lateral ventricles. Due to the sharp contrast between the CSF in the ventricles and surrounding tissue in T1-weighted images, measurement of ventricular volume is amenable to robust automatic segmentation. Thus, the ventricles are among the study focus in the research of brain tissue atrophy~\citep{nestor2008ventricular}. Particularly in AD research, ventricular enlargement, as a measurement of hemispheric atrophy rates, has been repeatedly reported as a marker of AD progression~\citep{nestor2008ventricular,fjell2009mini,kruthika2019multistage}.

\section{Discussion}
\label{sec:discussion}

In this study, we propose an $\ell_{1}$-type regularization for predictors following a hierarchical tree structure. Under the concept of a path diagram, the proposed penalty regulates the total effect of each predictor on the outcome. With regularity conditions, it is shown that under the proposed regularization, the estimators of the model coefficients are consistent in $\ell_{2}$-norm and the model selection is also consistent. By applying to a brain structural imaging dataset acquired from the ADNI study, the proposed approach identifies brain regions associated with a declination in memory. With regularization on the total effects, the findings suggest that the impact of atrophy on memory deficits is from small brain regions.

When the predictors follow a hierarchical structure, the ordinary lasso regularization may lead to biased results, as the predictors can be strongly correlated~\citep{zou2005regularization}. The proposed approach circumvents this issue by introducing the influence matrix as the penalty matrix in the regularization. We show that this is equivalent to applying the ordinary lasso regularization on the independent latent factors that generate the predictors. In this study, we focus on the estimation and interpretation of the model parameters and the total effects. An important follow-up question is to perform inference on the parameters and the total effects, which we leave to future work. Though this study is motivated by structural neuroimaging data, it can be generalized to any other area of research where hierarchical compositions are informative predictors.

\section*{Acknowledgments}


Data collection and sharing for this project was funded by the Alzheimer's Disease Neuroimaging Initiative (ADNI) (National Institutes of Health Grant U01 AG024904) and DOD ADNI (Department of Defense award number W81XWH-12-2-0012). ADNI is funded by the National Institute on Aging, the National Institute of Biomedical Imaging and Bioengineering, and through generous contributions from the following: AbbVie, Alzheimer's Association; Alzheimer's Drug Discovery Foundation; Araclon Biotech; BioClinica, Inc.; Biogen; Bristol-Myers Squibb Company; CereSpir, Inc.; Cogstate; Eisai Inc.; Elan Pharmaceuticals, Inc.; Eli Lilly and Company; EuroImmun; F. Hoffmann-La Roche Ltd and its affiliated company Genentech, Inc.; Fujirebio; GE Healthcare; IXICO Ltd.; Janssen Alzheimer Immunotherapy Research \& Development, LLC.; Johnson \& Johnson Pharmaceutical Research \& Development LLC.; Lumosity; Lundbeck; Merck \& Co., Inc.; Meso Scale Diagnostics, LLC.; NeuroRx Research; Neurotrack Technologies; Novartis Pharmaceuticals Corporation; Pfizer Inc.; Piramal Imaging; Servier; Takeda Pharmaceutical Company; and Transition Therapeutics. The Canadian Institutes of Health Research is providing funds to support ADNI clinical sites in Canada. Private sector contributions are facilitated by the Foundation for the National Institutes of Health (\url{www.fnih.org}). The grantee organization is the Northern California Institute for Research and Education, and the study is coordinated by the Alzheimer’s Therapeutic Research Institute at the University of Southern California. ADNI data are disseminated by the Laboratory for Neuro Imaging at the University of Southern California.


\appendix
\counterwithin{figure}{section}
\counterwithin{table}{section}
\counterwithin{equation}{section}
\counterwithin{lemma}{section}
\counterwithin{theorem}{section}



\section{Theory and Proof}
\label{appendix:sec:proof}

\subsection{Proof of Theorem~\ref{thm:consist}}
\label{appendix:sub:proof_thm_consist}

Before proving Theorem~\ref{thm:consist}, we first introduce the definition of $\kappa_{1},\kappa_{2}$ and $\kappa_{3}$, which are compatibility constants.
Let $\kappa_{\mathcal{R}}$ denote the \textit{compatibility constant} between $\mathcal{R}$ and the $\ell_{2}$-norm on $\mathcal{M}$,
  \[
    \kappa_{\mathcal{R}}=\sup_{\bbeta}\left\{\mathcal{R}(\bbeta):\bbeta\in\mathcal{B}_{2}\cap\mathcal{M}\right\},
  \]
where $\mathcal{B}_{2}=\{\bx\in\mathbb{R}^{p}:\|\bx\|_{2}\leq 1\}$ is the $\ell_{2}$-norm ball. Let $\varrho$ be some norm on $\mathcal{R}^{p}$ such that $\mathcal{R}(\bbeta)\leq\varrho(\bbeta)$ for any $\bbeta\in\mathbb{R}^{p}$. Use $\kappa_{\text{IC}}$ to denote the compatibility constant between the irrepresentable term and $\varrho$
  \[
    \kappa_{\text{IC}}=\sup_{\varrho(\bz)\leq 1}V\left[P_{\mathcal{M}^{\perp}}\left\{QP_{\mathcal{M}}(P_{\mathcal{M}}QP_{\mathcal{M}})^{-}P_{\mathcal{M}}\bz-\bz\right\}\right],
  \]
where $P_{\mathcal{M}}\bz$ denotes the \textit{projector} of $\bz$ on $\mathrm{span}(\mathcal{M})$. Theorem~\ref{thm:consist} is proved by first computing the constants $\kappa_{1},\kappa_{2},\kappa_{3}$.

\begin{proof}
  When $\alpha\neq0$, $\bD$ is invertible, thus $\tilde{\bD}$ has a nontrivial null space. Let $\varrho=\|\cdot\|_{1}$, the compatibility constants are computed as the following:
  \[
    \kappa_{1}=\kappa_{\text{IC}}=\|\tilde{\bD}_{\mathcal{S}^{c}}\bX^\top(\tilde{\bD}_{\mathcal{S}}\bX^\top)^{-}\mathrm{sign}(\bbeta^{*}_{\mathcal{S}})\|_{\infty},
  \]
  \[
    \kappa_{2}=\kappa_{\mathcal{R}}=\sup_{\bbeta}\left\{\|\tilde{\bD}\bbeta\|_{1}:\bbeta\in\mathcal{B}_{2}\cap\mathrm{span}(\tilde{\bD}^\top\mathcal{B}_{\infty,\mathcal{S}^{c}})^{\perp}\right\},
  \]
  \[
    \kappa_{3}=\kappa_{\varrho}=\sup_{\bbeta}\left\{\|\bbeta\|_{1}:\bbeta\in\mathcal{B}_{2}\cap\mathrm{span}(\tilde{\bD}^\top\mathcal{B}_{\infty,\mathcal{S}^{c}})^{\perp}\right\}.
  \]
  Since $\mathcal{R}$ and $\varrho$ are finite, $\kappa_{1},\kappa_{2},\kappa_{3}<\infty$. The rest of the proof can be found in \citet{lee2015model}.
\end{proof}

For the regularization $\mathcal{R}_{1}$, $\bD$ has a nontrivial null space since it is of full rank. To compute the compatibility constants, one can replace $\tilde{\bD}$ with $\bD$. The rest of the proof follows.


\section{Additional ADNI Results}
\label{appendix:sec:adni}

\subsection{Estimate}
\label{appendix:sub:adni_est}

Table~\ref{appendix:table:adni_est} presents the estimate of the direct and total effect of the relevant brain regions in Figure~\ref{fig:adni_brain}.
\begin{table}
  \caption{\label{appendix:table:adni_est}Estimated direct ($\hat{\beta}$) and total ($\hat{\gamma}$) effect of the relevant brain regions in Figure~\ref{fig:adni_brain}.}
  \begin{center}
    {\small
    \begin{tabular}{l c r r}
      \hline
      \multicolumn{1}{c}{ROI} & \multicolumn{1}{c}{Level} & \multicolumn{1}{c}{$\hat{\beta}$} & \multicolumn{1}{c}{$\hat{\gamma}$} \\
      \hline
      Intracranial & 0 & - & - \\
      Telencephalon (L) & 1 & - & - \\
      Telencephalon (R) & 1 & - & - \\
      CSF & 1 & 0.013 & - \\
      Cerebral Cortex (L) & 2 & - & - \\
      Cerebral Cortex (R) & 2 & -0.002 & - \\
      Cerebral Nucli (R) & 2 & -0.035 & - \\
      White Matter (L) & 2 & - & - \\
      White Matter (R) & 2 & - & - \\
      Ventricle & 2 & - & - \\
      Sulcus (L) & 2 & -0.034 & -0.046 \\
      Temporal (L) & 3 & - & - \\
      Limbic (L) & 3 & -0.006 & - \\
      Limbic (R) & 3 & 0.005 & 0.012 \\
      Occipital (L) & 3 & -0.022 & - \\
      Anterior WM (R) & 3 & - & - \\
      Inferior WM (L) & 3 & - & - \\
      Lateral Ventricle (L) & 3 & 0.009 & - \\
      Lateral Ventricle (R) & 3 & 0.006 & - \\
      Sylvian Fissure (L) & 3 & -0.046 & -0.046 \\
      MTG (L) & 4 & 0.001 & - \\
      Limbic (L) & 4 & 0.003 & 0.003 \\
      Limbic (R) & 4 & 0.037 & 0.037 \\
      IOG (L) & 4 & 0.140 & 0.140 \\
      Amygdala (L) & 4 & 0.067 & 0.067 \\
      Amygdala (R) & 4 & 0.032 & 0.032 \\
      Anterior DPWM (R) & 4 & - & - \\
      Inferior DPWM (L) & 4 & -0.014 & - \\
      Inferior Lateral Ventricle (L) & 4 & -0.122 & -0.122 \\
      Inferior Lateral Ventricle (R) & 4 & -0.064 & -0.064 \\
      MTG pole (L) & 5 & -0.006 & -0.006 \\
      IFO (L) & 5 & 0.129 & 0.129 \\
      SFO (R) & 5 & 0.014 & 0.014 \\
      \hline
    \end{tabular}
    }
  \end{center}
\end{table}

\subsection{Analysis results from competing methods}
\label{appendix:sub:adni_comp}

\begin{figure}
  \begin{center}
    \subfloat[Lasso ($\mathcal{R}_{2}$)]{\includegraphics[width=0.45\textwidth]{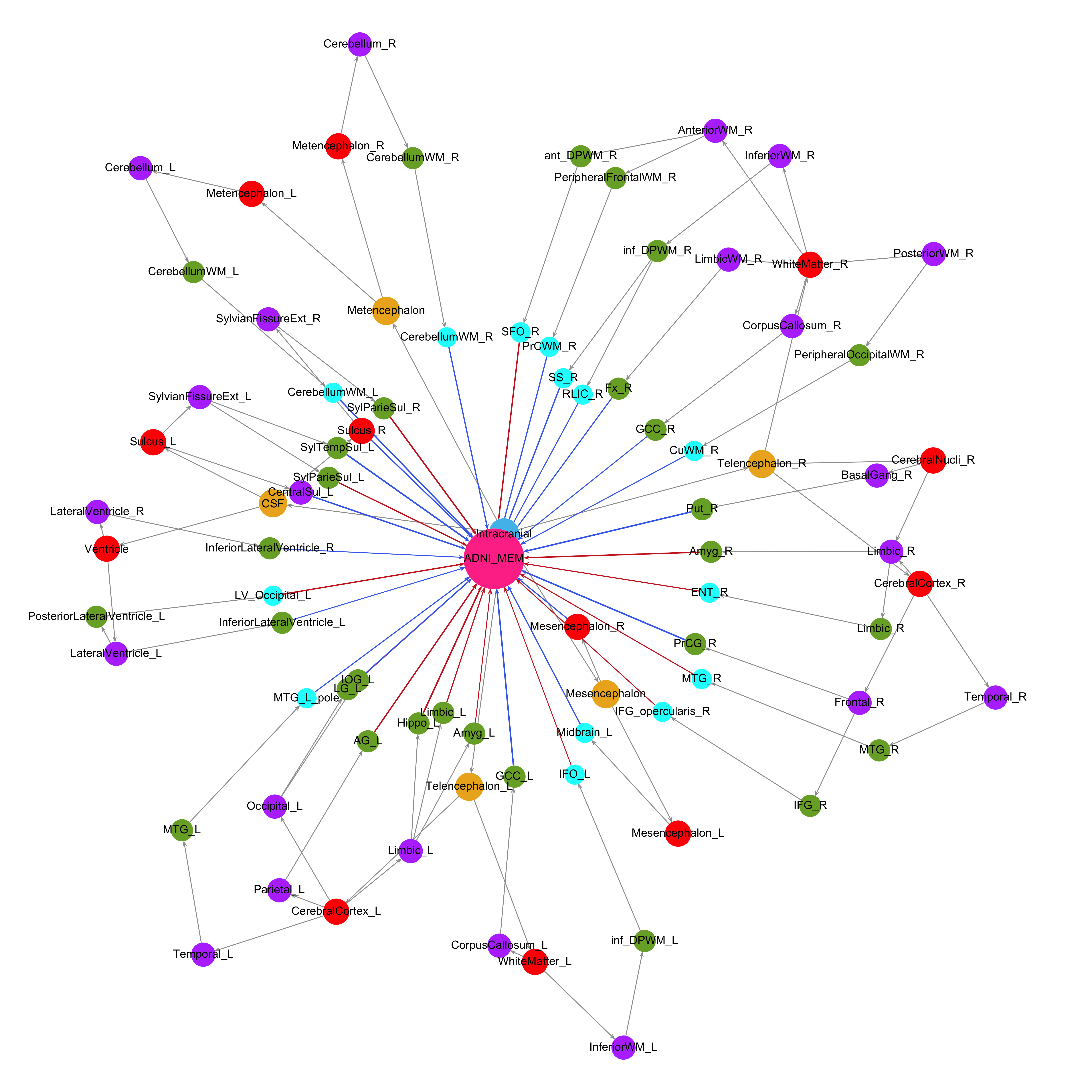}}
    \enskip{}
    \subfloat[Elastic net]{\includegraphics[width=0.45\textwidth]{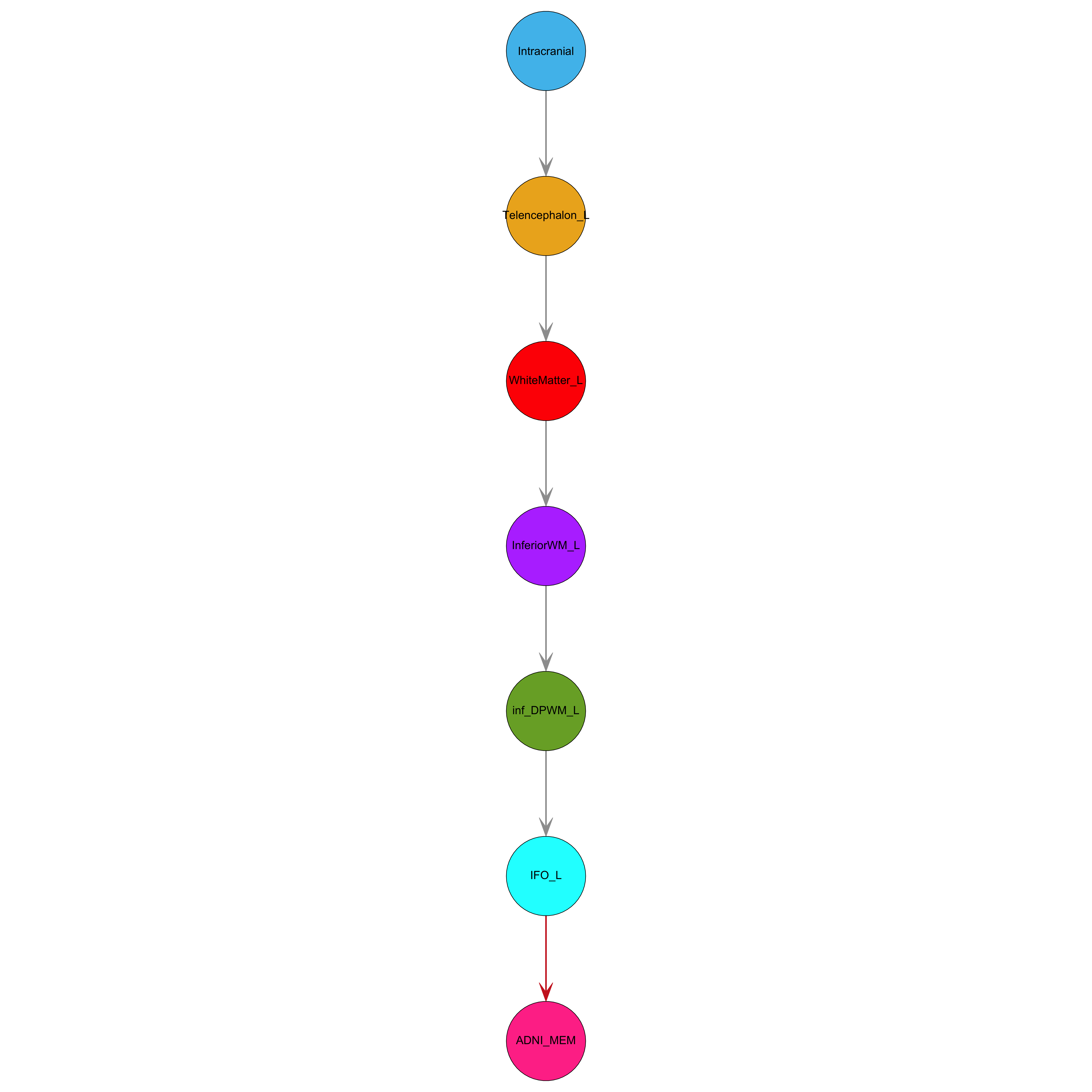}}

    \includegraphics[width=0.45\textwidth]{legend_level.png}
  \end{center}
  \caption{\label{appendix:fig:adni_comp_tree}The identified brain regions related to the ADNI\_MEM outcome using (a) the lasso ($\mathcal{R}_{2}$) and (b) the elastic net regularization putting into the tree diagram. Brain regions are colored by level. The gray arrows inform the hierarchical structure. A red arrow indicates a positive effect and a blue one indicates a negative effect. The width of the lines is proportional to the magnitude of the effect.}
\end{figure}

Figure~\ref{appendix:fig:adni_comp_tree} presents the identified brain regions that are related to the ADNI\_MEM outcome using the lasso ($\mathcal{R}_{2}$) and the elastic net regularization putting the regions into the brain segmentation tree diagram. From the figures, most of the regions identified by these two approaches are Level-4 and Level-5 regions, where the effect of some regions can be aggregated into the parent region in a higher level. Compared to the results in Figure~\ref{fig:adni_brain}, the hierarchical structure of the identified regions are less obvious as the information is ignored.

\bibliographystyle{apalike}
\bibliography{Bibliography}

\end{document}